\documentclass{ws-rv9x6}
\usepackage[square]{ws-rv-van}             % numbered citation/references (default)
\makeindex
\begin{document}

\setcounter{chapter}{1}
\chapter[EVOLUTION OF THE AGE STRUCTURED POPULATIONS ...]{EVOLUTION OF THE AGE STRUCTURED POPULATIONS AND DEMOGRAPHY\label{ch2}}

\author[A. {\L}aszkiewicz, P. Biecek, K. Bo{\'n}kowska, S. Cebrat]{Agnieszka {\L}aszkiewicz${}^2$,  Przemys{\l}aw Biecek${}^1$, \\ Katarzyna Bo{\'n}kowska${}^1$, Stanis{\l}aw Cebrat${}^1$}
%\index[aindx]{Author, F} % or \aindx{Author, F}
%\index[aindx]{Author, S} % or \aindx{Author, S}

\address{${}^1$ Department of Genomics, Faculty of Biotechnology, University of Wroc{\l}aw, ul. Przybyszewskiego 63/77, 51-148 Wroc{\l}aw, Poland. 
E-mail: cebrat@smorfland.uni.wroc.pl. \\
${}^2$ Institute of Immunology and Experimental Therapy, Polish Academy of Sciences, ul Weigla 12, Wroc{\l}aw, Poland}

\begin{abstract}
We describe the simulation method of modeling the population evolution using Monte Carlo based on the Penna model. Individuals in the populations are represented by their diploid genomes. Genes expressed after the minimum reproduction age are under a weaker selection pressure and accumulate more mutations than those expressed before the minimum reproduction age. The generated gradient of defective genes determines the ageing of individuals and age-structured populations are very similar to the natural, sexually reproducing populations. The genetic structure of a population depends on the way how the random death affects the population. The improvement of the medical care and healthier life styles are responsible for the increasing of the life expectancy of humans during the last century. Introducing a noise into the relations between the genotype, phenotype, and environment, it is possible to simulate some other effects, like the role of immunological systems and a mother care. One of the most interesting results was the evolution of sex chromosomes. Placing the male sex determinants on one chromosome of a pair of sex chromosomes is enough to condemn it for shrinking if the population is panmictic\index{panmictic} (random-mating is assumed). If males are indispensable for taking care of their offspring and have to be faithful to their females, the male sex chromosome does not shrink. 
\end{abstract}

\body

\section{Introduction}\label{sec1.1}
Let's start from the end - the end of our life, when our strength weakens and finally the death takes its toll. The mysterious nature of this phenomenon eludes scientific investigations, leaving it still the ,,unsolved biological problem'' \cite{MEDAWAR}. Most of the ageing theories try to describe mechanisms which are suspected to shape the slow decay of life. They invoke various forms of damage to DNA, cells, tissues and organs. The very popular free radical theory of aging may serve as an example \cite{HARMAN56,Harman81}. However, evolutionary theories try to reach the roots of the phenomenon, seeking to explain why we age; in other words: how is it possible that senescence, which decreases the fitness of the individual, has evolved at all? Nevertheless, the proper use of these theories requires the awareness of their limitations, all these concepts should be ,,handled with care'', as it was suggested by Le Bourg \cite{LE BOURG98,LE BOURG99}.

\begin{figure}
\includegraphics[height=0.95\textwidth,angle=270]{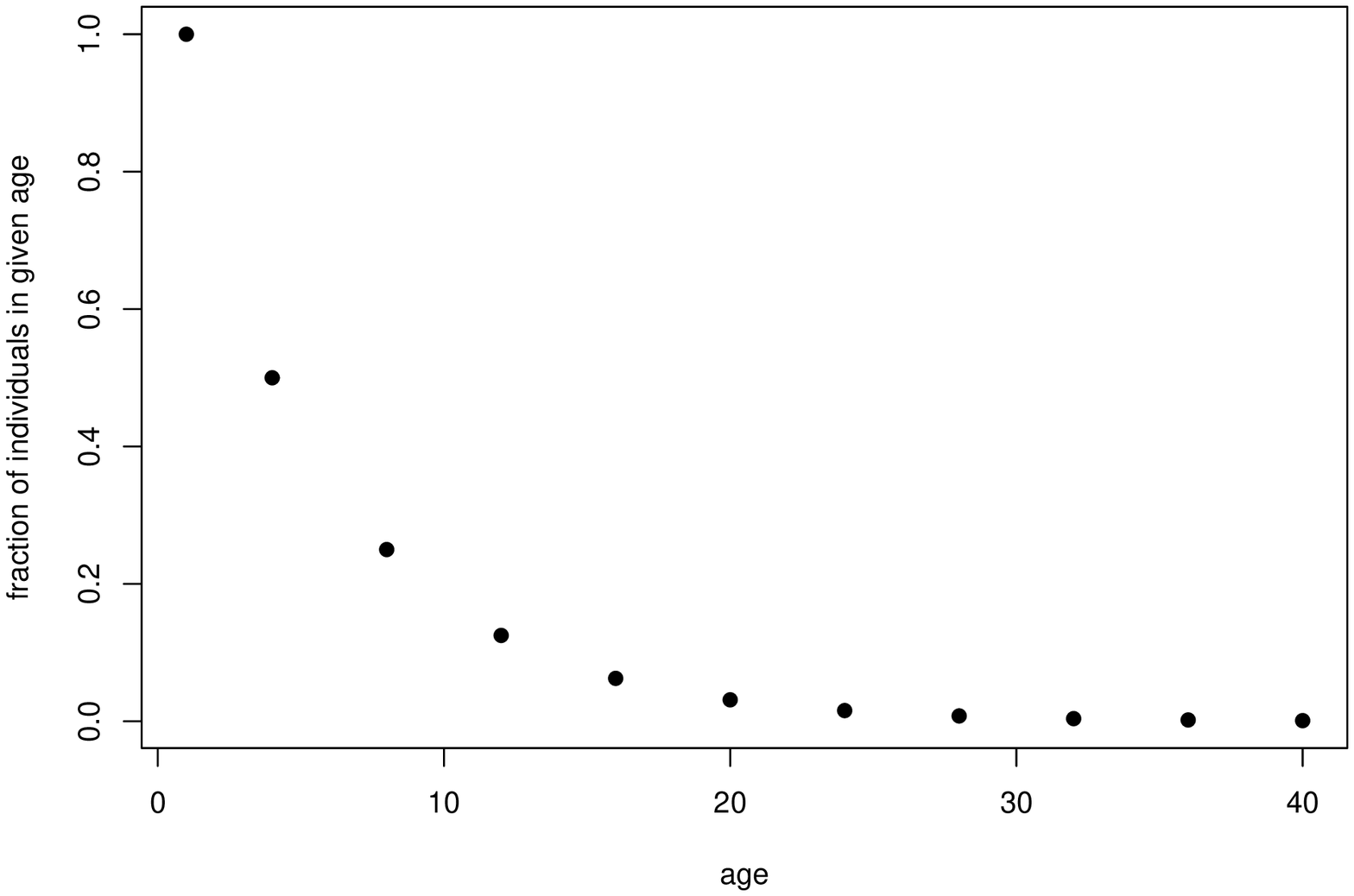}
\caption{\label{fig1}Distribution of number of test tubes of a given age \cite{MEDAWAR}. It is equivalent to the age distribution in populations where individuals do not age and die only because of random accidents.}

\includegraphics[height=0.95\textwidth,angle=270]{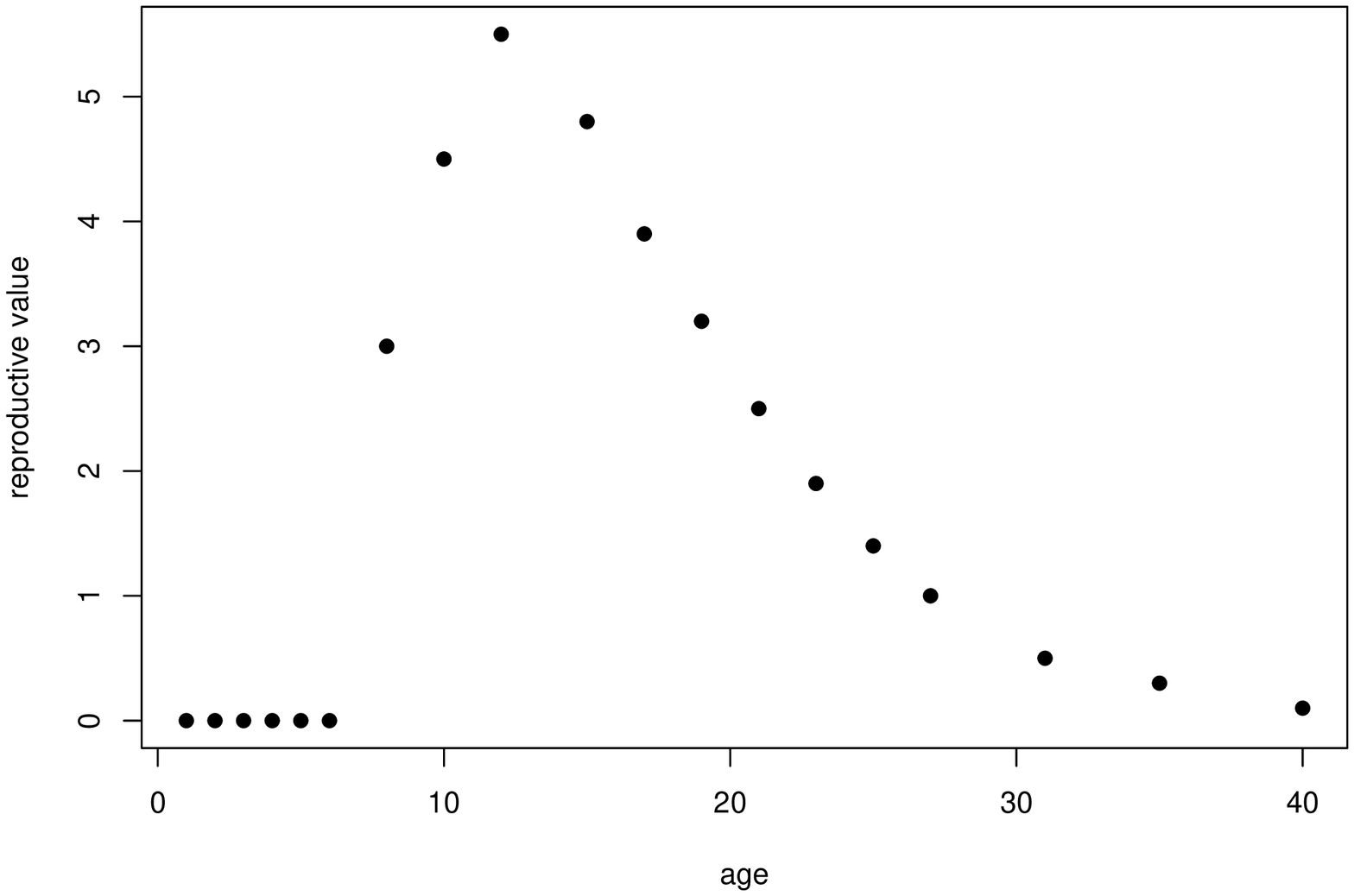}
\caption{\label{fig2}Reproduction values of different age groups according to Goldsmith \cite{GOLDSMITH04}. The youngest organisms, before the minimum reproduction age do not contribute to the reproduction. The contribution of the older groups is lower not because they are less fertile but because there are less of them in population.}
\end{figure}

One of the first biologists who proposed that ageing is a product of evolutionary forces was Weismann \cite{Weismann}. He believed that there is a genetically programmed death which eliminates older members of population, providing more resources for the younger generation. In this context, ageing represents an adaptation - advantageous for the species, even if it has a negative effect on the individual fitness. Although Weismann changed his views on the evolution of ageing over the course of his life, in the literature his name is associated with the first version of his programmed death theory.
In response to this initial idea, Medawar presented his own theory of ageing \index{theory of ageing}during an inaugural lecture at University College London in 1951 \cite{MEDAWAR}. He emphasised that the programmed death theory is logically circular; Why do animals age and die? Because they become worn out and decrepit and consequently valueless for the species. But why are they worn out and decrepit? Because they age. In other words, we are not ageless because we age. 
Medawar took another approach; the existence of a post-reproductive period is one of the consequences of senescence; it is not its cause. Developing his model of immortal species, Medawar used as a metaphor random breaking and replacing of test tubes. In a chemical laboratory, equipped with a stock of the test tubes, the broken tubes are replaced by the new ones (age = 0), after some time the number of test tubes of a given age will decline exponentially with age (Fig. \ref{fig1}). The older the test tubes are, the fewer there will be of them - not because they become more vulnerable with increasing age, but simply because the older test-tubes have been exposed the more times to the hazard of being broken \cite{MEDAWAR}.

Avoiding Weismann's trap,\index{Weismann's trap} Medawar assumed the equal probability of reproduction for young and old individuals. Since a non-ageing individual tends to have statistically the same number of progeny, every year, from puberty to death, the total number of produced progeny increased linearly with age. 

These two functions: exponential decrease of the number of individuals with age and linear increase of progeny with age shape the contribution of each age group to the total reproduction potential of population. As it can be seen in Fig. \ref{fig2}, the reproductive value of the group of older individuals is low. In Medawar's words: ,,The older the age group, the smaller is its overall reproductive value... This is not because (the test tubes) of the senior group are individually less fertile but merely because there are fewer of them; and there are fewer of them not because they have become more fragile - their vulnerability is likewise unaltered - but simply because, being older, they have been exposed more often to the hazard of being broken'' \cite{MEDAWAR}.

Summing up, Medawar's idea states that the population loses individuals with time because of the random death, so the reproductive effect of the group of older individuals decreases with age, therefore the selection for their continued survival and reproduction should also weaken. Since these verbal arguments have been formalized,  the idea began life as a mathematically coherent theory\cite{HAMILTON66,CHARLESWORTH80}. It turns out that before the onset of reproduction, the force of natural selection is highest and constant. After this time, however, the force of natural selection progressively falls, reaching zero around the cessation of reproduction. As a consequence, the genes beneficial early in life are favoured by natural selection while genes which should be expressed and necessary for surviving the late ages, after the reproduction period, can freely accumulate deleterious mutations. 

According to mutation\index{mutation} accumulation theory proposed by Medawar himself, mutational pressure introduces new hereditary factors, which could lower the fertility or viability of the organisms. However, if they are expressed late enough, the force of selection will be too attenuated to oppose their establishment and spread \cite{MEDAWAR}. In this so called ,,selection shadow'' many germ-line mutations neutral for young individuals, but with late deleterious effects accumulate passively over generations leading to senescence and death \cite{KIRKWOOD00}.
According to presented theory, ageing is a nonadaptive trait, being only a by-product of the way natural selection operates. It claims that senescence evolved because it is out of reach of natural selection and there was no possibility to eliminate it. That is why these theories are sometimes called the law of unintended consequences. However, the scientific opposition believe in ,,Death by Design'' and try to revive Weisman's idea of ageing evolving as an adaptation. They criticise Medawar's model of non-ageing species, which in their opinion is oversimplified and ignores the evolutionary importance of older animals. They claim that the probability of reproduction of old non-ageing individuals is not lower than that of young ones (Weisman's trap), nor is it equal (Medawar's idea); in fact the old immortal individual has a higher probability of reproduction, since it is more mature and better fit. Moreover, older non-ageing animals have lower probability of death than younger less experienced individuals: ,,The old king is less likely to die in the war than the young foot soldier'' \cite{GOLDSMITH04}. Using this kind of arguments ,,opposition'' is trying to discredit Medawar's assumptions and call reliability of his model into question. 

Finally, what is certain about the end of our biological existence? The end is certain.

\section{The Penna Model}
\subsection{Description of the standard Penna model}
In this chapter we will present the results of computer simulations of populations' evolution based on the Penna model\cite{Penna}. Hundreds of papers using this model or citing it have been already published. Authors of these papers (including Penna himself) claim that it is an ageing model and very often it is suggested that the model assumes the Medawar hypothesis of ageing \cite{Moss}. In fact, the Penna model does not assume the Medawar hypothesis but the results of simulations using this model support the Medawar hypothesis. Moreover, the Penna model could be very useful in describing many phenomena connected with the evolution of biological populations, like altruism \cite{cebrat}, menopause \cite{suzana}, the role of the immunological system \cite{biecek}, of mother care \cite{MARF}, even the evolution of sex chromosomes \index{sex chromosomes}\cite{sex}. There are two main versions of the model, haploid and diploid.\index{haploid}\index{diploid} We will describe the diploid version only, because it fits much better to the biological reality. In the description of the model we will try to introduce some genetic terminology just to show the correspondence between some genetic elements and mechanisms and the virtual world of the model. It would facilitate further contacts of non-biologist readers with biologists. 

In the model, populations are composed of individuals, each one represented by its genome\index{genome} composed of two haplotypes which are the bitstrings~$L$~bits long. Bits represent genes. Bits placed in the corresponding position \index{locus (loci)}(locus/loci) in the bitstrings are called alleles.\index{bitstring}\index{allele} If a bit is set to 0, it represents a wild form of an allele (a correct one), if a bit is set to 1, it represents the defective allele. If one allele is defective and the other one in the corresponding locus of the second bitstring is wild, then two situations could be considered: if the defect is dominant\index{dominant} - the function (or phenotype\index{phenotype}) is defective, if the defective \index{allele}allele is recessive\index{recessive} - the function of the locus is normal, if both bits at the same positions are defective, the phenotype is always defective. In all presented here simulations we have assumed that all defective alleles are recessive. The main assumption of the Penna model is the chronological switching on the pairs of alleles; in the first Monte Carlo step (MCs), the alleles at the first locus are switched on, in the second step, the alleles at the second locus and so on. One MCs corresponds to a time unit/period. The number of defective loci which kill the individual is one of the model parameter called the threshold $T$. Death of individual caused by the expression of the $T$ phenotypic defects is called the genetic death. If the individual expresses all pairs of alleles in its two bitstrings, then even if it does not reach the threshold $T$ it is going to die because of reaching the ,,absolute maximum age'' $L$. In fact, when parameters are properly set, individuals do not reach that age, they are dying earlier because of the genetic death. 

When an individual reaches the minimum reproduction age $R$ (after switching on $R$ pairs of alleles) it can reproduce. In all our simulations we assume the conditions of sexual dimorphism and we declare that the sex of a newborn can be set as a male or a female with an equal probability. A~female at the reproduction age produces a gamete.\index{gamete} To do that the diploid genome of that female is copied and during this process a new mutation\index{mutation} is introduced into each haplotype \index{haplotype}(a bitstring) at the randomly chosen position with a probability $M/L$ per loci (thus $M$ is the average number of mutations per haplotype). If the bit chosen for mutation has a value 0 - it is changed for 1, if it is already 1 it stays 1 - there are no reversions.\index{reversion} The two bitstrings recombine in the process mimicking the crossover\index{crossover} with a probability $C$. The position, where the two bitstrings are cut for the crossover is a randomly chosen. One of the two products of the crossover is randomly chosen as a female gamete and then, the male individual at the reproduction age is chosen randomly from the whole pool of ,,adult'' males in the population. This male produces its gamete in the same way as a female has done it. 
The female and male gametes form the diploid genome of an offspring, its sex - male/female is set with an equal probability. In the next MCs its first locus will be checked for the genetic status of its bits (genes). In one MCs an age of all individuals increases by one and each female at the reproduction age can give a birth to a new baby with a probability $B$ or can give a birth of $B$ new babies. There is another mechanism controlling the birth rate in the model. It is a logistic equation of Verhulst: $V=1-N_t/N_{max},$ where $N_t$ is the number of individuals in the population in the $t$-step of simulation, $N_{max}$  is a maximal size of the population called also a capacity of the environment and $V$ is a probability that a new offspring will survive its birth. 

In fact there are only seven parameters in the standard Penna model:
\begin{itemize}
\item	$L$ - the number of loci,
\item	$M$ - the average number of mutations introduced into the haploid genome during the gamete production (usually $M$=1 per haploid genome per generation),
\item	$B$ - the number of offspring produced by each female at reproduction age at each time step or the probability of giving the offspring,
\item	$R$ - minimum reproduction age,
\item	$T$ - the upper limit of expressed defects, at which an individual dies,
\item	$C$ - the probability of a crossover between parental haplotypes during the gamete production, 
\item	$V$ - the Verhulst factor $V=1-N_t/N_{max}$.\index{Verhulst factor}
\end{itemize}

\subsection{Results of standard simulations}
\begin{figure}
\includegraphics[height=0.95\textwidth,angle=270]{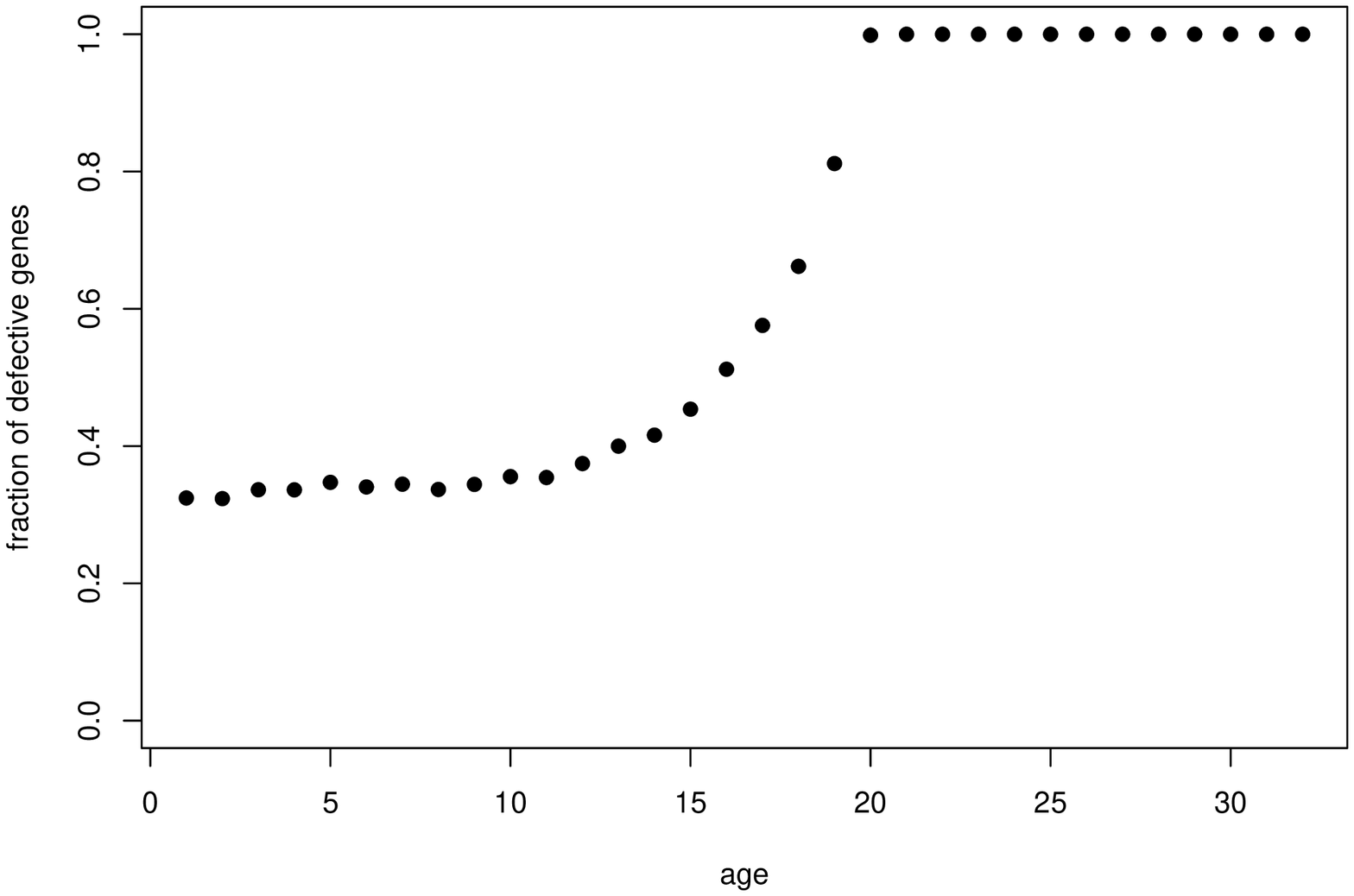}
\includegraphics[height=0.95\textwidth,angle=270]{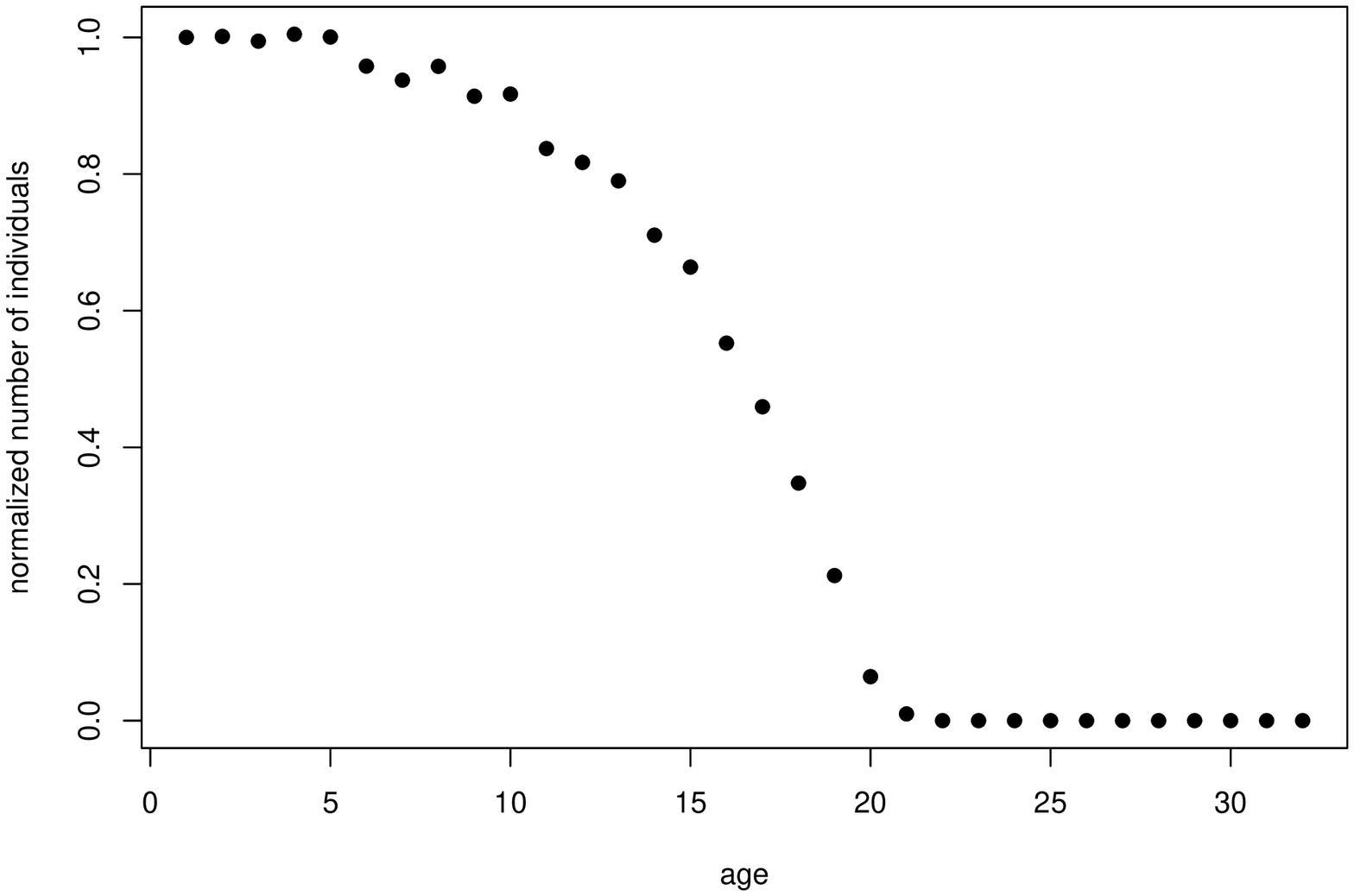}
\caption{\label{fig3}Distribution of defective genes (bits set to 1) in the genetic pool of virtual Penna populations - top panel, and age distribution of the corresponding population - lower panel.
The age corresponds to the number of MC steps and to the number of bits switched on. Parameters of simulations: $L$=32, $M$=1, $C$=1, $B$=1, $T$=3, $R$=8.  }
\end{figure}

There are many possibilities of starting the simulations. Usually we start with randomly generated population of half of $N_{max}$ size with perfect genomes (all bits set for 0). Randomly generated means here evenly distributed age of individuals and equal numbers of females and males. During the simulation, a mutational pressure introduces mutations into haplotypes while a selection tends to eliminate them. Figure \ref{fig3} shows the distribution of defective genes in the genomes in equilibrium. Among the genes expressed before the minimum reproduction age ($R$) the fraction of defective alleles is the lowest. After $R$ the fraction of defective genes grows with the age of their expression, eventually reaching 1 for the late ages which means that there are no functional genes in the whole genetic pool\index{genetic pool} which could be necessary for surviving during those late periods of life. Such a distribution of defective genes determines the age structure of populations. The mortality of the youngest individuals is the lowest and it is growing with age. The position of the first genes which are set in the whole population for 1 determines the maximum life span of individuals in the population. More precisely, the maximum life expectancy is ($T$-$1$) steps longer than indicated by the position of the first locus set to 1.

Note that in these populations individuals can reproduce with equal probability to the end of life. Thus, the results support the Medawar hypothesis of ageing assuming that even if individuals can reproduce to the end of their life, their impact on the total reproduction potential of population decreases and selection doesn't care about their genetic status allowing the accumulation of mutations in genes expressed during the late periods of life.

\subsection{The role of parameters in modelling the age structured populations}
\subsubsection{Random death}
\begin{figure}
\includegraphics[height=0.95\textwidth,angle=270]{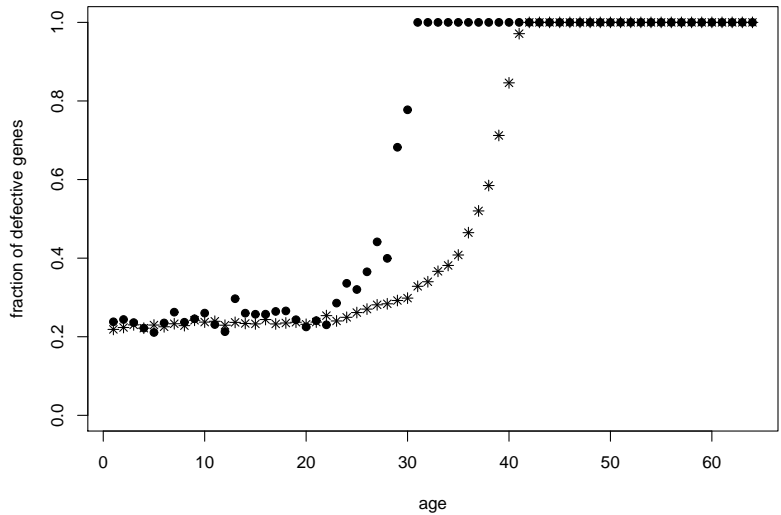}
\caption{\label{fig4}The difference in distribution of defective genes in the genetic pool of populations evolving under Verhulst factor operating at birth only (stars) or "killing" all individual independently of their age (circles).}
\end{figure}

One of the most important factors in the population simulations is the method of keeping the stable size of populations. There are two factors in the Penna model: birth rate $B$ and Verhulst factor $V$ responsible for that. The birth rate has to be set high enough to secure at least the replacement of fraction of dying individuals in the evolving populations. A higher birth rate leads to the roughly exponential growth of population and that is why the logistic equation of Verhulst is introduced. Deaths due to Verhulst factor may be introduced in a few different ways and it is very important if this factor kills the individuals randomly during their lifespan or at birth only \cite{SaMartins,Bonkowska}. We set the Verhulst factor only for newborns  which means that each newborn is additionally tested by Verhulst factor for surviving. Figure \ref{fig4} shows the difference in distribution of defective genes in the genetic pool of populations evolving with differently operating Verhulst factors. 

One can wonder why the random death could affect the populations so differently. There are two reasons. One is that random death killing at birth kills individuals independently of their genetic status. If it kills individuals during their lifespan then, the older individuals have to pass the tests of their genomes several times. These older organisms have larger fraction of their genomes correct which has been testified directly by their age. The other conclusion is that the random death weakens the selection for genetic status of individuals. In fact, in Nature such a random death operates mainly at very early stages of development or at very late periods of life.

\subsubsection{Threshold T}
The average life expectancy, which supposes to mirror the population's health, rose dramatically during the 20th century. The most striking example is East Asia, where the life expectancy at birth increased from less than 45 years in 1950 to more than 72 years nowadays (it means that it was increasing by six months each year!). So far, however, Japanese women enjoy the longest expectation of life: it approached 85 years in 2003, in comparison to 63 years in 1950 and 44 years in 1900 \cite{Kinsella05}. This outstanding achievement of our civilization becomes a challenge. Combined with the decrease in fecundity (number of children born by one woman), it has brought the global ageing of the human population and fears of pension crises. Such deep demographic changes strongly influence the economic, educational and social policies. Therefore, there is an urgent need for correct forecasts of expectation of life for future decades. Unfortunately, there is no agreement among scientists in these predictions.

\textbf{How much can the human life span be extended?}

This question could have many contexts. It can be understood more philosophically as the quest for the truth about the possibility of physical immortality. In other words: whether we are biologically designed to die and nothing can be done about it or rather, putting stop to ageing is only the matter of time and eventually human beings will live forever, unless killed by random death. More practical, however, this is a question about the demographic trends in the coming decades, and there are at least two schools which aspire to answer it. They are characterized by different points of view on current human populations' demographic trends and predict different future demographic scenarios. 

\textbf{Adding life to years...;} The first idea, developed by Fries \cite{FRIES80}, is rooted in the conviction that there is an inherent limit imposed on the human longevity by biology, which determines the maximal life span. Consequently, there must also be a ceiling to human life expectancy. Fries even assessed the position and shape of the ,,ideally rectangular survival curve'', which would be the result of the hypothetical elimination of all premature deaths, and set ultimate life expectancy level at 85 years. The intuitive consequence of this process of rectangularization\index{rectangularization} should be the so called compression of morbidity. If the onset of infirmity is postponed (e.g. by changes in lifestyle or improvements in medicine), the period with the debilitating effects of illness or disability reduces, because of the fixed maximum age. This concept predicts that increasing life expectancy should not result in the extension of morbidity: in ,,ever older, ever more feeble, and ever more expensive-to-care-for populace'' \cite{FRIES80}, but rather in the compression of infirmity into a short period prior to death, associated with the reduction in the scale of health care systems.

Summing up, the presented interpretation of demographic changes may be illustrated using a popular catch-phrase as a process of ,,adding life to years''. There is no place for prophets of physical immortality in this scenario.

\textbf{...or years to life?} More and more scientists come to believe that this theory does not properly describe the demographic reality. Oeppen and Vaupel \cite{Oeppen02} presented a comparison between the forecasts done in the past and real-life development as a story about breaking all the limits predicted by demographers to life expectancy: from 64.75 years calculated as its ultimate level by Louis Dublin in 1928 to 85 years set again by Olshansky et al. in 1990 \cite{Olshan}. The main message of their article is that there are no signs that life expectancy is approaching any limit and that the belief in such limits led to underestimation of life expectancy in the past and definitely will not help in forecasting the future trends. This observation does not discredit the rectangularization hypothesis itself. As long as the increase in maximum age at death is slower than the increase in average age at death (life expectancy) the survival curve will show rectangularization. But such rectangularization without the belief in the fixed maximum age will not provide us with such a simple and precise answer to what we can expect in the close future as it was in Fries' scenario.

And reality seems to be even more complicated and difficult to forecast. As a matter of fact, the pattern of changes in the survival curve that occurred in the first half of 20th century did show rectangularization. But scientists report that this trend has been replaced in some countries by a near parallel shift of the curve to the right \cite{YASHIN01}. This leaves the future wide open for speculations. 

More generally, some scientists stress the significance of medical interventions and public health as the determinants of mortality changes; others see their origin mostly in economical, social, cultural and behavioural factors. Nevertheless, one century is a too short  period for any significant reconstruction of the genetic pool\index{genetic pool} of the human population, and thus in broad terms the observed mortality transition must have been driven by the changes in the individual/environment relation and thus by the increase in ,,threshold value T'', whatever the detailed scenario of this process was. Consequently, it should be possible to simulate the changes in the human population just by increasing the individuals' tolerance to the number of expressed defective characters.

Since the $T$ parameter in the Penna model describes the individuals' tolerance to the deleterious traits that they suffer from, we used it to model the dynamic of demographic changes \cite{Laszkiewicz}. When $T$ is set for one, it means that every single deleterious phenotypic trait kills the individual. The higher the $T$ value means that  each individual copes better with its genetic load. In the real world, this value might correspond to the standard of medical care, lifestyle, diet, hygiene, education and all other factors that help an individual to escape some fatal effects of his genetic traits. Therefore, this is a very capacious parameter and it can be generally interpreted as the factor in charge of the relation between individuals and environment. Notice that changes in the age structure of human populations were observed in very short period - in Japan it was only 50 years - less than two generations. There was no time for rebuilding the genetic pool of populations. Thus, all changes suppose to correspond to relations between the environment and populations. That is why we used for these studies the populations in the equilibrium and then we have changed the values of T parameter \cite{Laszkiewicz}. 

\begin{figure}
\centering
\includegraphics[height=0.9\textwidth,angle=270]{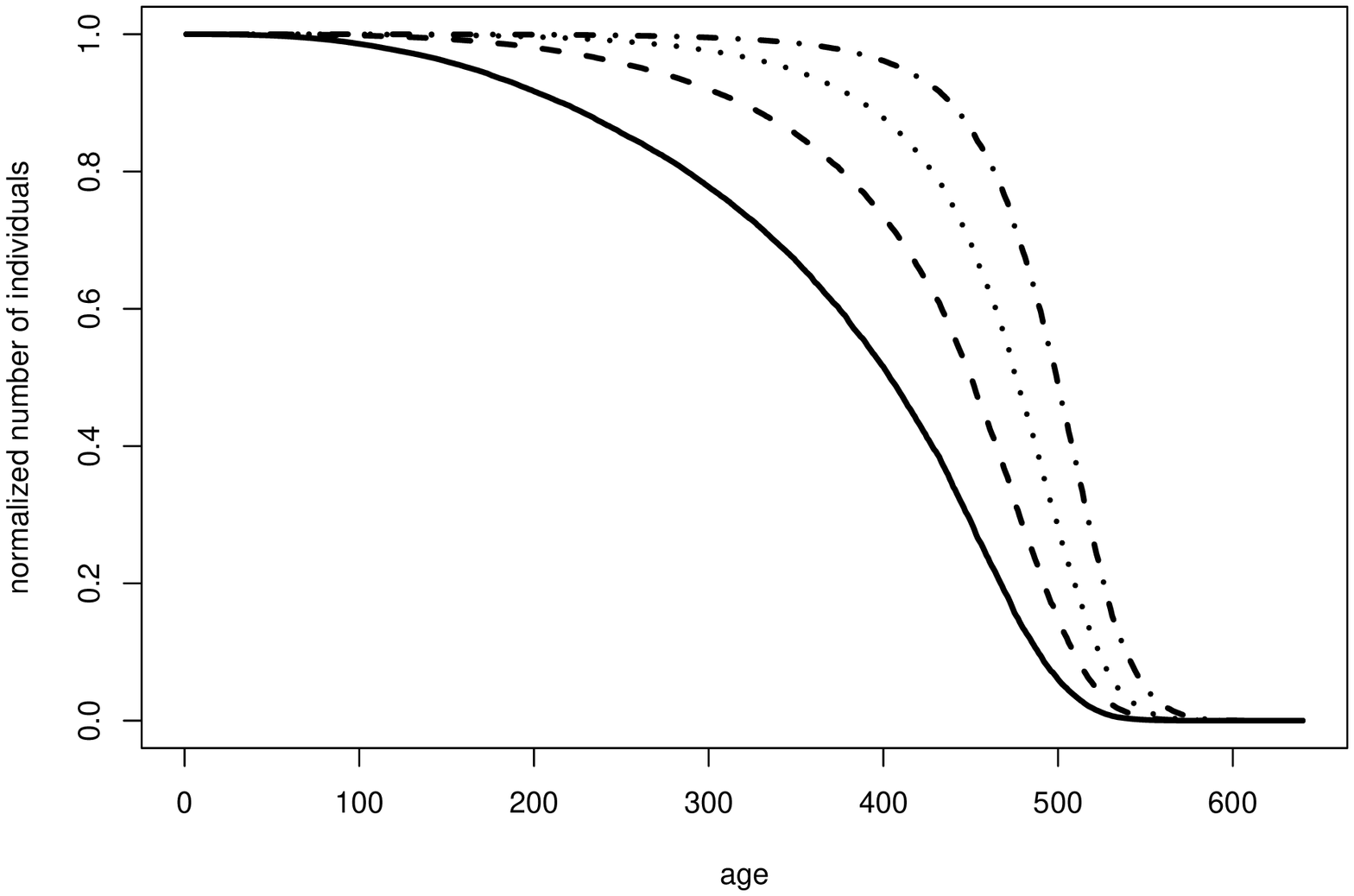}
\caption{\label{fig5}Effect of changes of the T parameter values on the age distribution of populations.  Parameters of simulations of the original population: L=640, R=200, M=1, C=1, B=0.25, T=3. After 80 000 MCs under threshold T=3 the anticipated age structure of such population under higher T value was calculated; T=3, T=4, T=5, T=6.3 from lower left to upper right curves. For details see the text.}

\includegraphics[height=0.9\textwidth,angle=270]{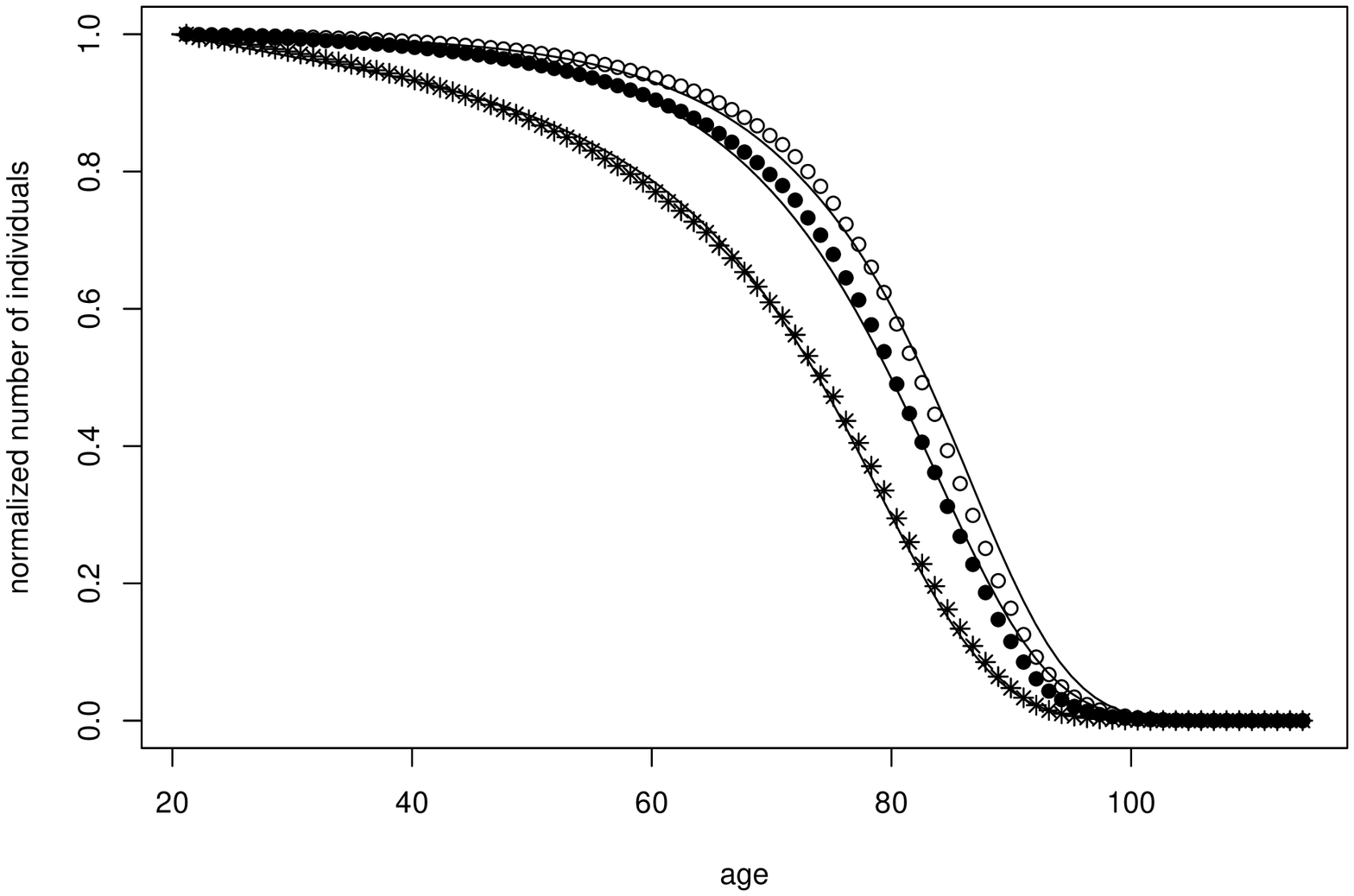}
\caption{\label{fig6}The age distribution of the Swedish populations in different periods (1930-39, 1980-89 and 2000-2003 - lines  from lower left to upper right) and corresponding results of simulations under different T values: 3.8, 4,9 and 5.4 symbols from lower left to upper right. For more details about rescaling the x-axis values see the text.
}
\end{figure}

The plots in Fig. \ref{fig5} show the role of changes of the $T$ parameter values. In these series of simulations populations evolved under the values of parameters as follows: $L$ = 640, $R$ = 200, $B$ = 0.25, $N_{max}$ = 50 000, $C$ = 1. Populations were allowed to evolve for 80 000 MC steps under the threshold $T$ = 3. Then, one million zygotes were produced and the life span of each zygote was anticipated without further evolution, just on the basis of their genetic load. This was repeated for different values of parameter $T$. When these values were not integers, the probability of surviving under a given $T$ was equal to a fractional part of $T$, which means that individuals under threshold $T$ = 3.3 had a probability 0.7 to die when three bad mutations were reached. In this way the anticipated age distributions of individuals for different thresholds $T$ were prepared and plotted in Fig. \ref{fig5}. The increase of the threshold $T$ value leads to the rectangularization of the curve describing the age distribution of the population with relatively small shift of the maximum age to the right. Comparison of these results with the data from the real Swedish population shows that the increase in the $T$ value can mimic the demographic changes in the age distribution of the human population. Data presented in Fig. \ref{fig6} are based on those presented in Fig. \ref{fig5} but the x-axis has been rescaled to fit the human life span; the period from 20 - 100 years in the human life span corresponds to the period from minimum reproduction age $R$ = 200 to the maximum age in simulations. Plots generated by simulations fit very well the age distribution of human populations of second half of the 19th century and part of the 20th century \cite{Laszkiewicz}. However, for the end of the 20th century and for the latest data rectangularization shown by simulations seems to be too strong. The model renders very well the level of mortality in the middle ages but the mortality of the oldest individuals is overestimated. Interestingly, demographers claim that life expectancy at older ages relatively stable so far has increased at a spectacular pace just since the second  half of the 20th century. This observation leads to the hot debate about the tail of the human mortality curve.

Probably the mortality curve would be better represented if the gradient of the number of expressed genes per year was introduced. This means that a higher number of genes would be expressed in younger ages and progressively this number would decrease in older ages. Such a modification seems to be biologically legitimated and it would be very interesting to test it by simulations in the future. An interesting continuation of this work is the modelling of the increase in the $T$ value coupled with the decrease in parameter $B$. Such an approach was used by to study the effects of the dramatic decrease in the birth rate observed in the developed countries on the age structure of the human population \cite{Bonkowska1}.

Summing up, increase in parameter $T$ value leads to the rectangularization of the survival curve with relatively small increase in the maximal age, reflecting very well these general demographic trends observed in the human populations. However, how it influences future mortality trends remains a matter of speculations.

\textbf{Shift of the minimum reproduction age}

\begin{figure}[h!]
\includegraphics[height=0.95\textwidth,angle=270]{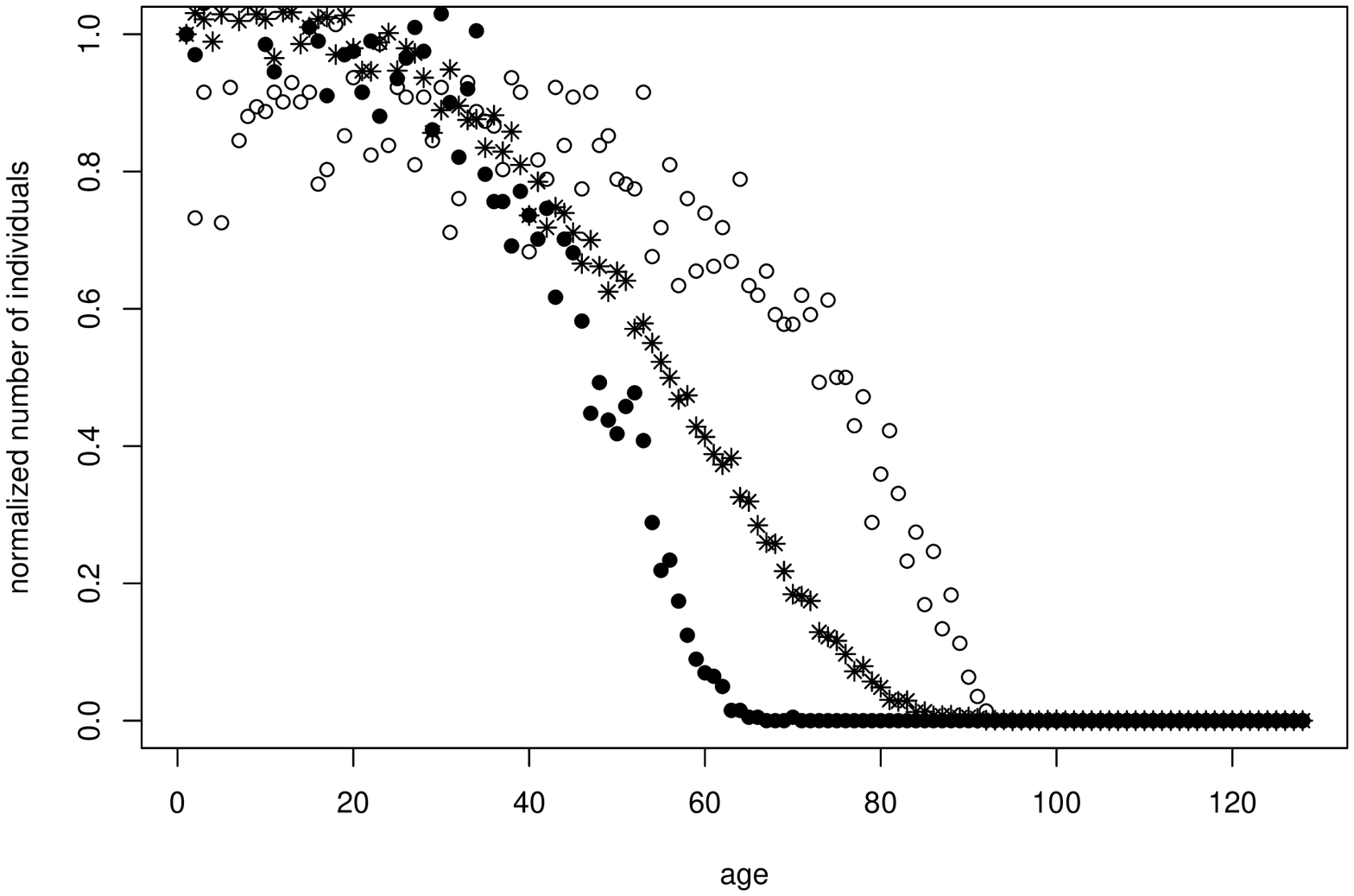}
\caption{\label{fig7}The effect of the increased minimum reproduction age on the age distribution of population. Populations evolved under different $R$ parameter values: 5 - filled circles, 20 - stars,  50 - open circles. The rest of parameters: $L$ = 128, $M$ = 1, $C$ = 1, $B$ = 1, $T$ = 3.}

\includegraphics[height=0.95\textwidth,angle=270]{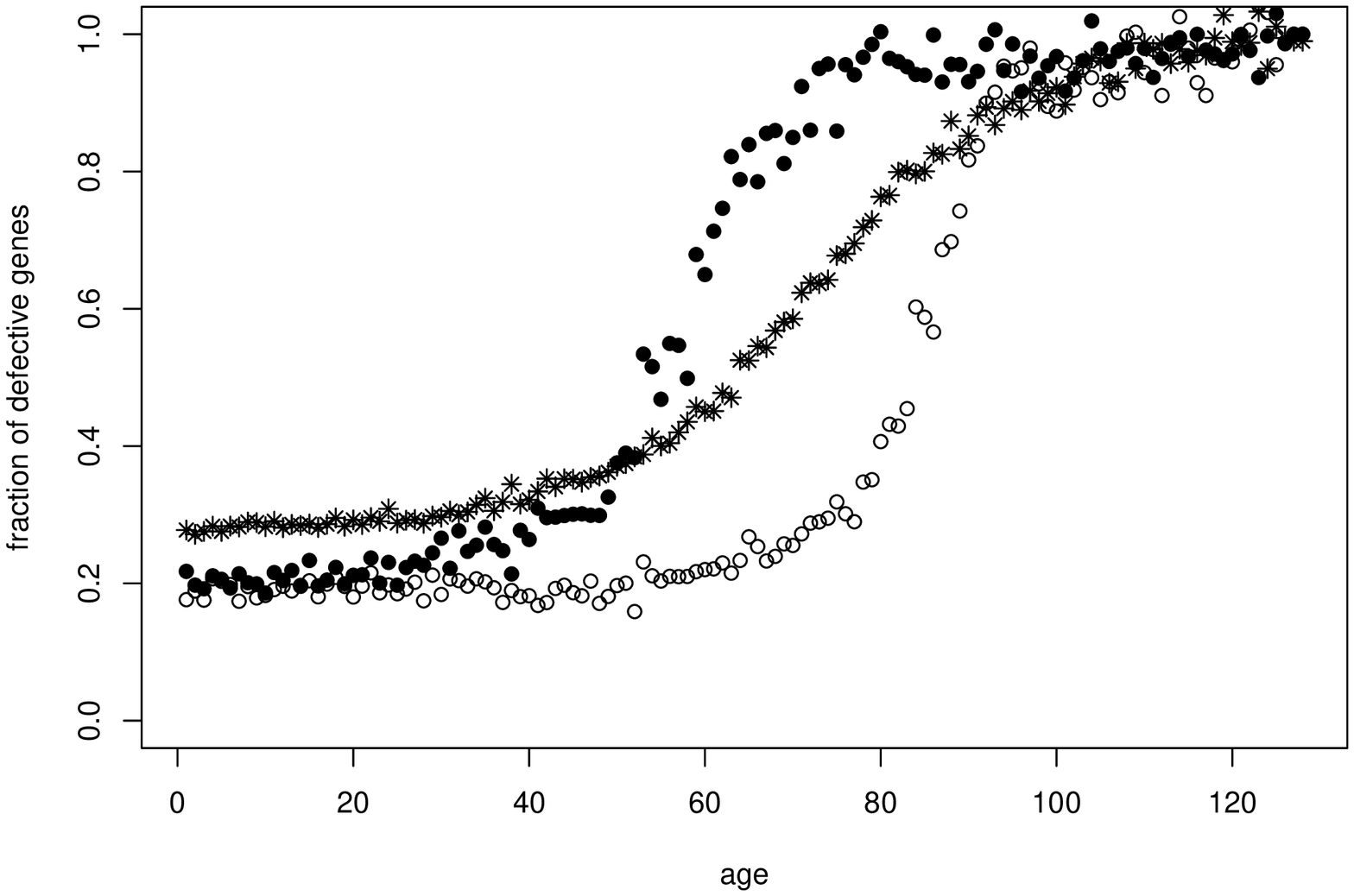}
\caption{\label{fig7b}The effect of the increased minimum reproduction age on the distribution of defects. Populations evolved under different $R$ parameter values: 5 - filled circles, 20 - stars,  50 - open circles. The rest of parameters: $L$ = 128, $M$ = 1, $C$ = 1, $B$ = 1, $T$ = 3.}
\end{figure}

\begin{figure}[h!t]
\includegraphics[height=0.95\textwidth,angle=270]{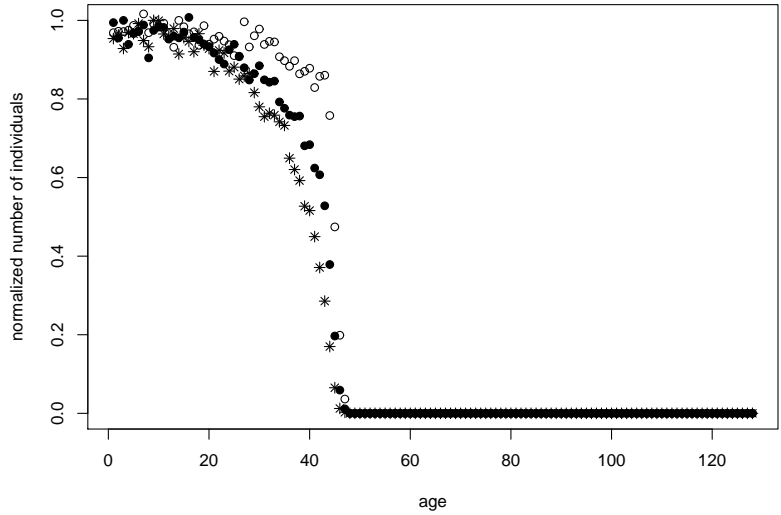}
\caption{\label{fig7c}The effect of evolution of population forced to change the age of reproduction to 10 - stars, 20 - full circles, 40 - open circles, 50 - population extinct. }
\end{figure}

Some peoples suggest that the best way to increase the human life span is to shift the minimum reproduction age for the later periods of life. Of course it is a good idea, which is shown in Fig. \ref{fig7}. When the simulations are started with the increased reproduction age - $R$ the population shifts the whole reproduction period to the older ages and the increase in the life expectancy corresponds roughly to the shift of $R$ and shift in the distribution of defective genes (Fig. \ref{fig7b}). Nevertheless, one has to remember that in this series of simulations populations evolved under higher $R$, thus evolution adapt these populations to such conditions where the longer time for reaching the puberty has been advantageous. The situation is not as optimistic, when we check the populations which already evolved under lower minimum reproduction age and then they were forced to shift the reproduction to the later ages. When the population which evolved under $R$ = 5 was forced to start the reproduction after the age of 50 - it was extinct (Fig. \ref{fig7c}) \cite{PhDthesis}.

\textbf{Penna with noise}

One can obtain some, rather unnatural, results of simulations using the Penna model; 
\begin{itemize}
\item 	The simultaneous death (or exactly the same lifespan) of individuals possessing exactly the same genomes i.e. in clones, in highly inbreed populations or twins, which obviously is not true \cite{Pletcher}. In Nature, for example in the inbreed lines of mice we still observe some distribution of the life spans of individuals - the life span of individuals is not precisely determined by their genomes. 
\item 	In the standard Penna model the threshold $T$ of defective phenotypic traits determines the age of ,,genetic death''.\index{genetic death} That means that at the age $< T$ no one individual can die because of its genetic status. In Nature a higher mortality of newborns is observed.
\end{itemize}

To overcome these problems with the Penna model we have introduced the noise to the model.

\clearpage
 
 \section{The noisy Penna model}
For the ,,noisy extension'' of the model, the diploid sexual Penna version has been used. In our version of the standard model, the Verhulst factor controls the birth-rate and there are no random deaths of organisms later during the lifespan. Individuals die only because of the genetic death when they reach the threshold $T$ of the expressed defective phenotypes. In the noisy version of the model, there is no declared threshold $T$. Instead, we have introduced the fluctuations of the state of organisms. The variance of fluctuations increases with the number of switched on defective loci \cite{biecek}. 

The state of an individual $i$ is denoted $I)i(t) \in \mathcal R$ is defined as a composition of the inner state\index{inner state} of the individual  (denoted $P_i(t) \in \mathcal R$) and a state of environment  (denoted $E(t) \in \mathcal R$). Thus
\begin{equation}
I_i(t) = E(t) + P_i(t),
\end{equation}
where 
$E(t) \sim \mathcal N(\mu_{E(t)},\sigma^2_E)$ corresponds to the fluctuations of environment 
in time $t$ while $P_i(t) \sim \mathcal N(\mu_{P_i(t)},\sigma^2_i(t))$ corresponds to the inner fluctuations of the individual $i$ in time $t$.
In the simplest case, the expected value of both fluctuations is $\mu_{P_i(t)} = \mu_{E(t)} = 0$ and the variance of the state of an individual depends on its number of defective loci $g_i(t)$ expressed till time $t$, i.e.
\begin{equation}
\sigma^2_i(t) = \sigma^2_0 + g_i(t)\sigma_d^2.
\end{equation}
One may modify $\mu_{E(t)}$ and include a drift or seasonal changes in the environment.\index{environment}

\begin{figure}
	\centering
		\includegraphics[height=0.95\textwidth,angle=270]{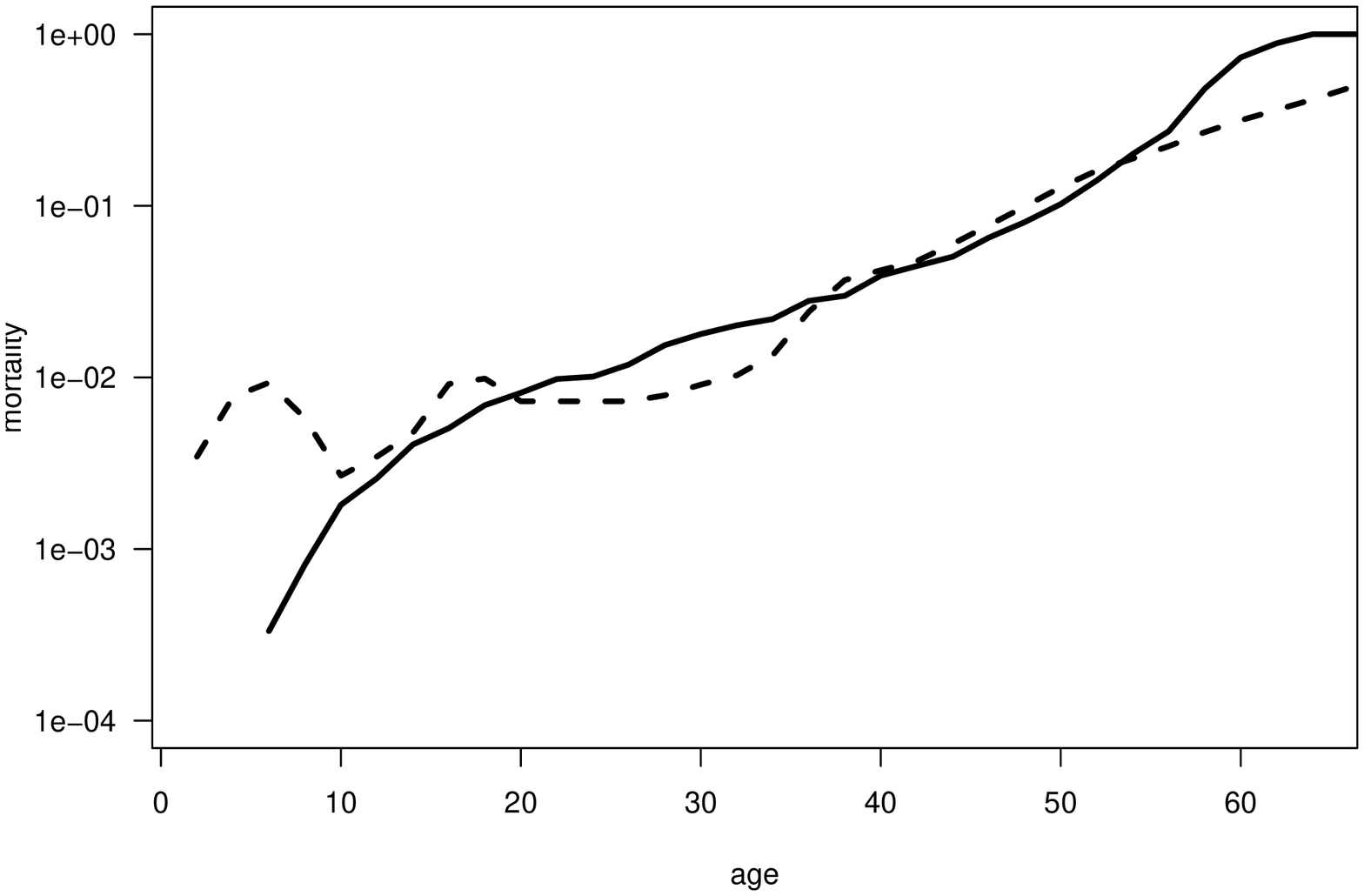}
		\includegraphics[height=0.95\textwidth,angle=270]{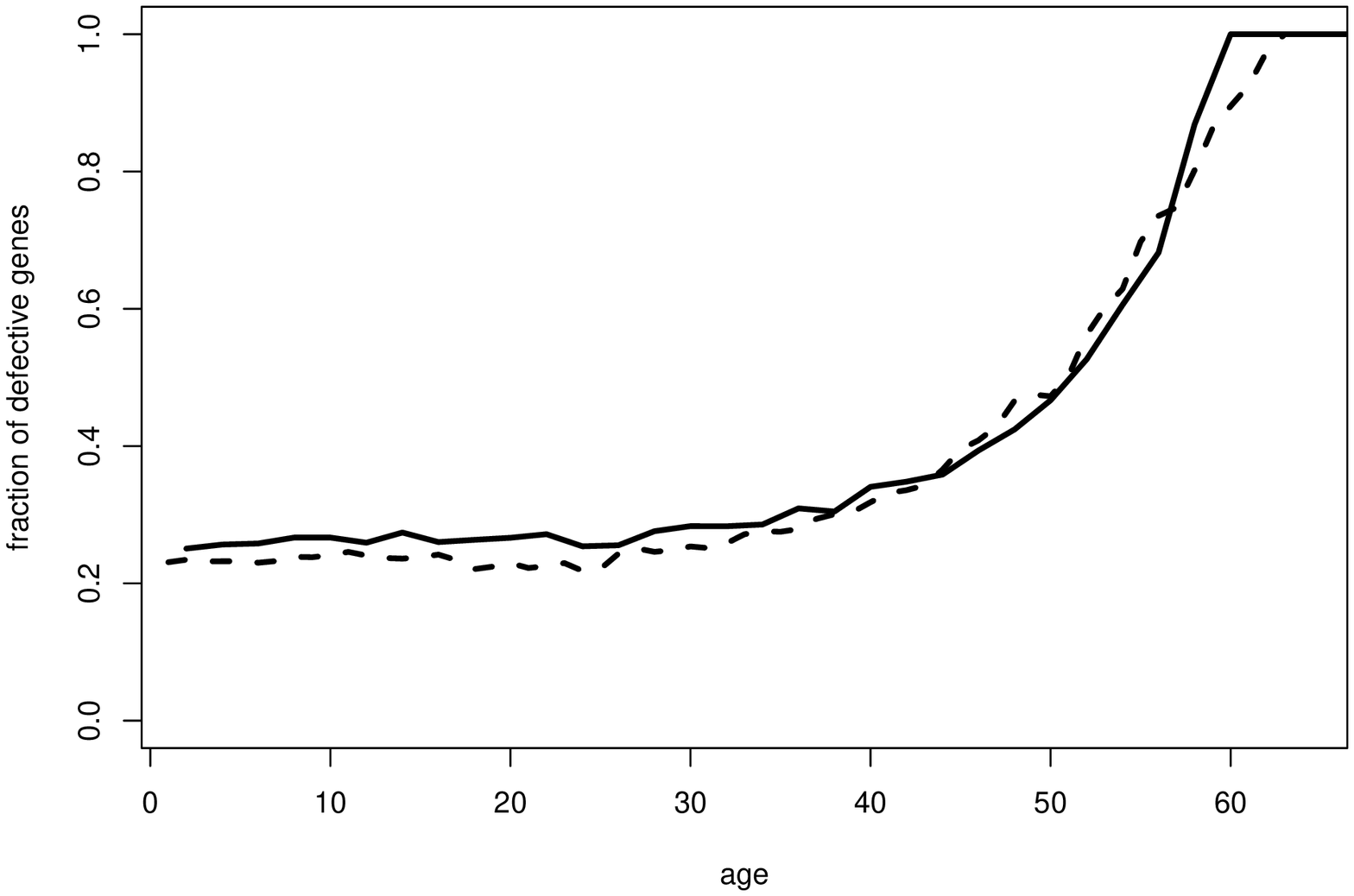}
		\caption{\label{fig8}The mortality curve (upper plot) and frequency of defective genes (lower plot) in the standard Penna model (solid line) and in the noisy Penna model (dashed line). Note the logarithmic y-axis of the mortality curve, x-axis is scaled in the arbitrary age units.}
	\label{fig:penna}
\end{figure}

Both the standard Penna model and the model with the noise produce very similar results. In both cases we observe characteristic distribution of defective genes expressed after the minimum reproduction age and a very low mortality of the youngest individuals. The largest difference between the two models concerns the mortality of individuals during the first two time units. For threshold $T=3$ in the standard model, there are no genetic deaths during the first two time units. In the noisy model, organisms may die even before the expression of any defect because of fluctuations (see Fig. \ref{fig8}). % and \ref{fig9}).

%
%\begin{figure}
%	\centering
%		\includegraphics[height=0.95\textwidth,angle=270]{figs/pennaStandard.eps}
%		\includegraphics[height=0.95\textwidth,angle=270]{figs/defektyPennaStandard.eps}
%		\caption{\label{fig8}The mortality curve (upper plot) and frequency of defective genes (lower plot) in the standard Penna model. Note the logarithmic y-axis of the mortality curve, x-axis is scaled in the arbitrary age units.}
%	\label{fig:penna}
%\end{figure}
%
%\begin{figure}
%	\centering
%		\includegraphics[height=0.95\textwidth,angle=270]{figs/pennaNoise.eps}
%		\includegraphics[height=0.95\textwidth,angle=270]{figs/defektyPennaNoise.eps}
%		\caption{\label{fig9}The mortality curve (upper plot) and frequency of defective genes (lower plot) in the noisy Penna model.}
%	\label{fig:defekty}
%\end{figure}

\section{Mother care}

In the noisy Penna model, the state of an environment affects all individuals regardless of their age but it is  possible to introduce some biologically legitimate bias in the relations individual - environment.
For example, the mother care as some kind of protection of the babies against the influence of fluctuations of the environment during the first periods of their lives affects these relations significantly. 
To model the mother care, the effect of the states of the environment on the newborns is decreased by (see \cite{MARF} for more details)
$$
\rho(i,t) = 1 - exp(-(age(i,t)+1)/\lambda_{MC}).
$$
The individual dies if $\rho(i,t) E(t) + P_i(t) > F$.
After $\lambda_{MC}$ steps the effect of the mother care is negligible while in the early stages of life it is significant. 

We call this ,,mother care''\index{mother care} to stress that this effect influences the very first periods of life but it could be the proper feeding of newborns with mother's milk as well as an intensive neonatal medical care.
In Fig. \ref{fig10} we present results for $\lambda_{MC}=4$. The distribution of defects is similar for both models and the mortality curve differs only in early stages of life.

\section{Adaptation to the environmental conditions - learning}

In the noisy Penna model, the variance of fluctuations of an individual state is a sum of its inner noise and the noise of an environment. The impact of these two components on the state and evolution of population is different. The personal component is independent for each individual while the environmental noise is the same for each individual. That is the reason why the reactions of individuals for the environmental fluctuations are diversified. 
In the version of the noisy Penna model presented above, the fluctuations have Gaussian distribution with the average $\mu_{E(t)}$. Now, we introduce a signal into the expected state of the environment. The signal $\mu_{E(t)}$ is a periodical function with the period D, thus $\mu_{E(t)} = \mu_{E(t+D)}$. Individuals know that the signal is periodical, and are equipped with a mechanism of learning the signal. They estimate components of the signal by weighted average of the state of the environment in survived periods. In the more formal way, the learning mechanism affects the expected state of individual fluctuations
$$
\mu_{P_i}(t) = \sum_{j=1}^{\infty} L(i,t-j*D) w_j E(t- j*D)
$$
where $L(i,t) = 1$ if individual $i$ lived at time $t$ and $0$ otherwise while weights $w_j$ are 
$$
w_j = e^{-(j-1)/\lambda} - e^{-j/\lambda}.
$$

\begin{figure}[t!h]
	\centering
		\includegraphics[height=0.95\textwidth,angle=270]{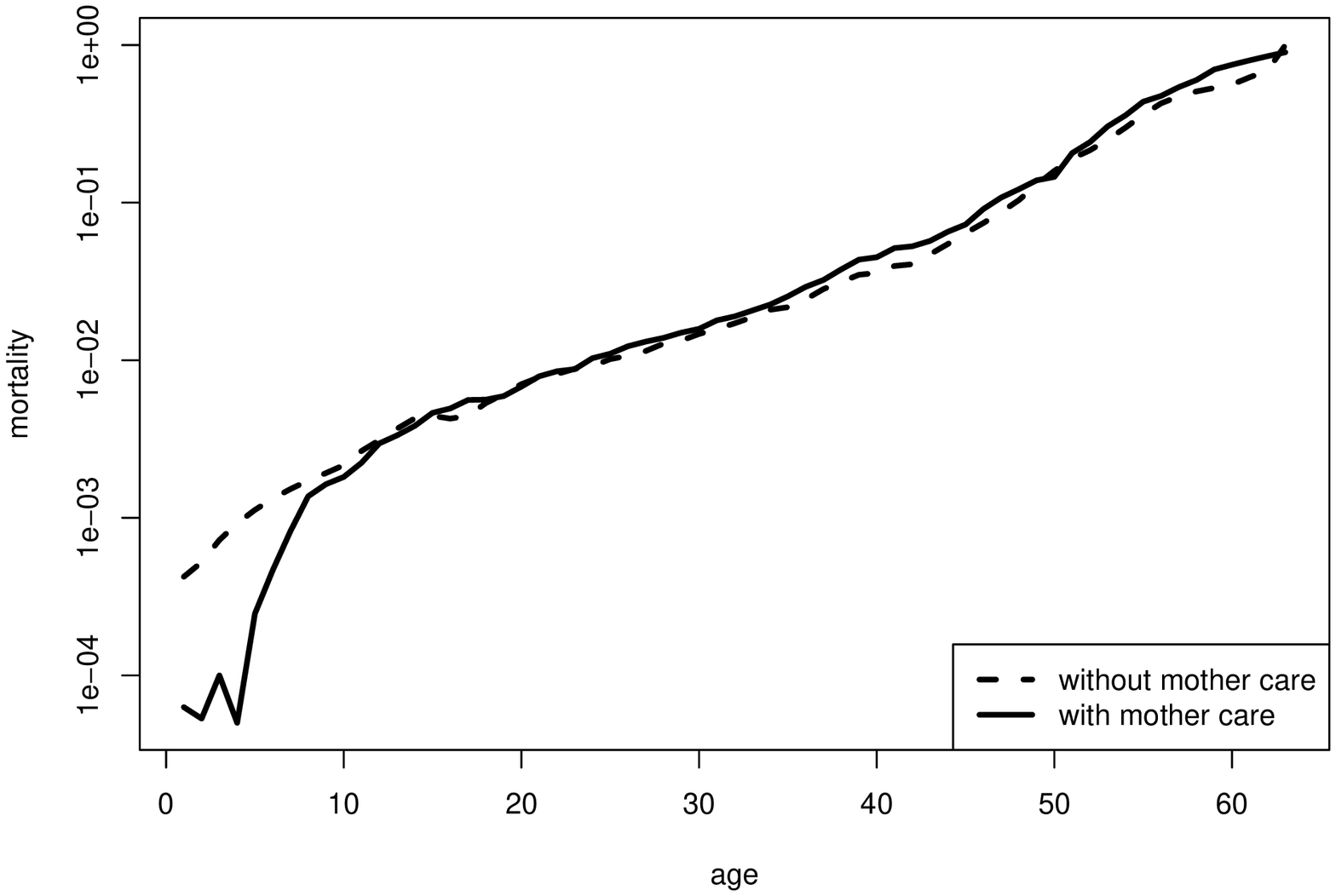}
		\caption{\label{fig10}Mortality curves for the population with and without mother care. There is no learning mechanism, thus in early years the mother care protects newborns and after this period the mortality with and without mother care is similar.}
%\end{figure}
%
%
%\begin{figure}[t!h]
	\centering
		\includegraphics[height=0.95\textwidth,angle=270]{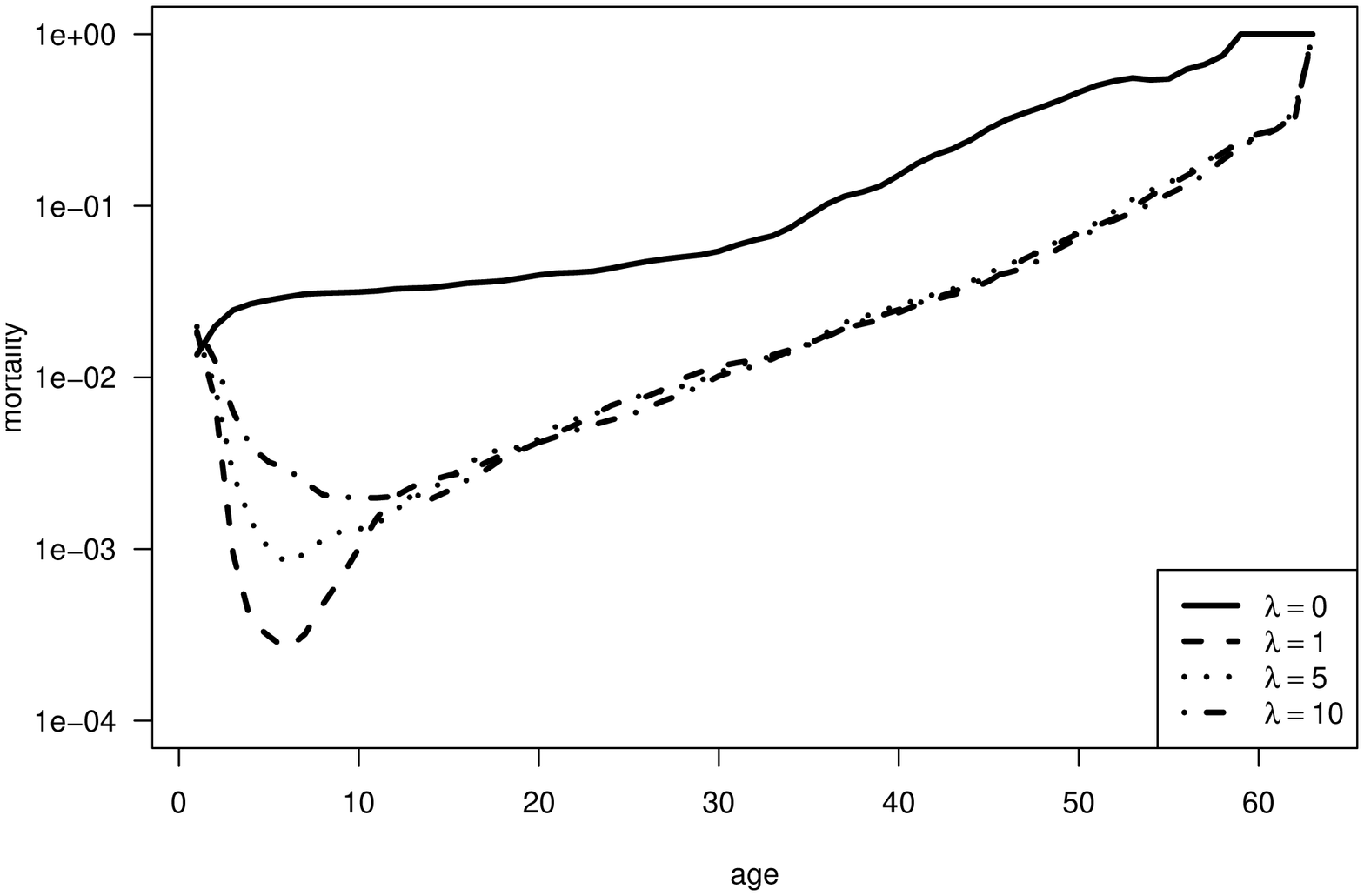}
		\caption{\label{fig11}The mortality curves for different learning coefficients $\lambda$.}
\end{figure}

This adaptation mechanism allows to reduce the mortality in case when an individual have learned the periodical signal. Results for different $\lambda$ are presented in Fig. \ref{fig11}. 
It is also observed in real populations that mortality of newborns is higher than of a bit older individuals. 
The results depend on the maximal signal value $\mu_{E(t)}$ and do not depend on the form of the periodic function, thus results for constant $\mu_{E(t)}=A$ are similar to those obtained with $\mu_{E(t)}=A \sin(t/\pi)$ (results not shown).

The next question is how the intensive protection of newborns against environment fluctuations could influence their mortality during the later periods of life. It is rather well known effect that children who are very strongly protected against any infections during the first periods of life and live in almost sterile conditions are more vulnerable for infections later. 

\begin{figure}[h!]
	\centering
		\includegraphics[height=0.95\textwidth,angle=270]{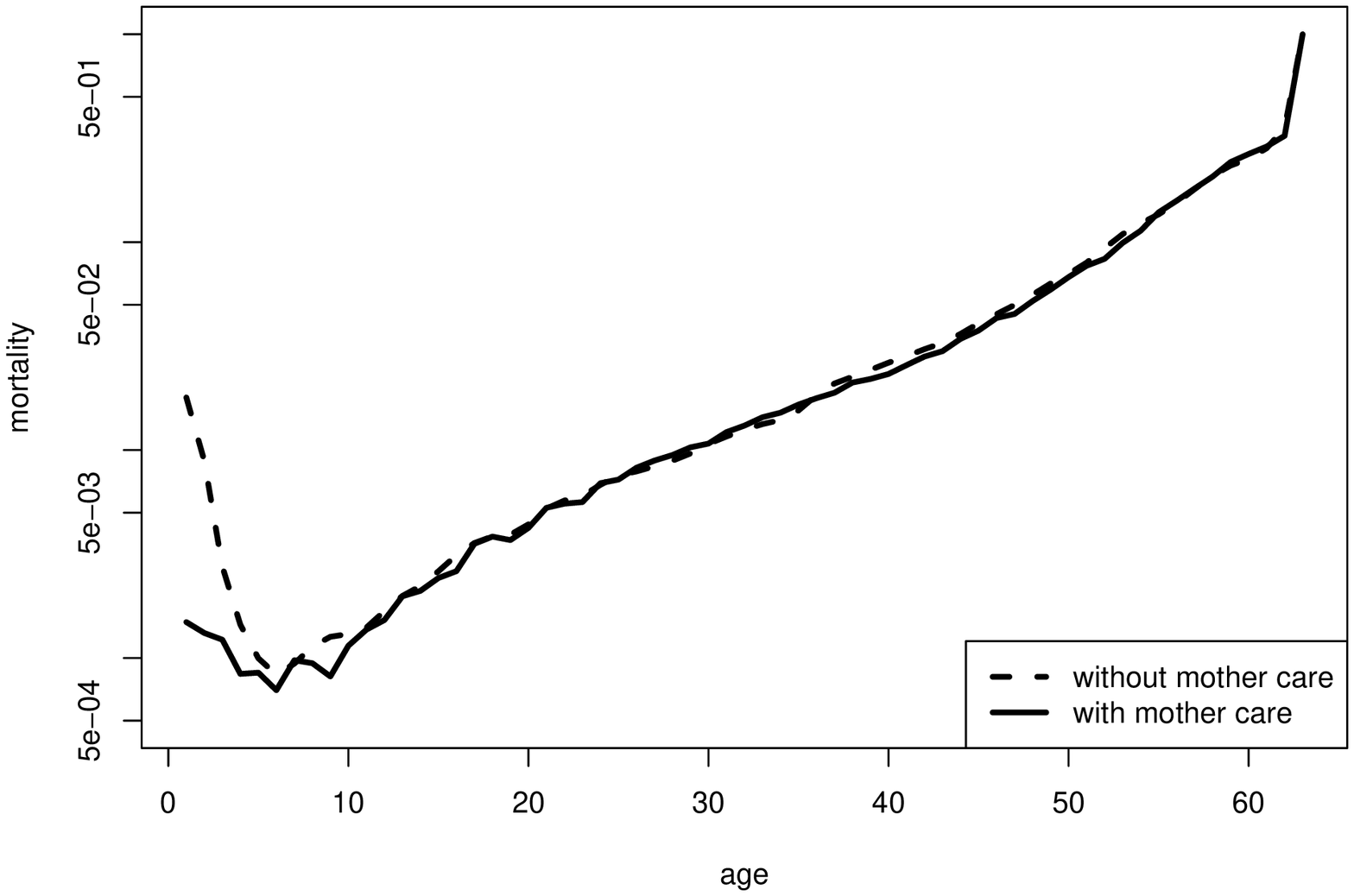}
		\caption{\label{fig12}The mortality curve for learning coefficient $\lambda=5$ with and without mother care. }

		\includegraphics[height=0.95\textwidth,angle=270]{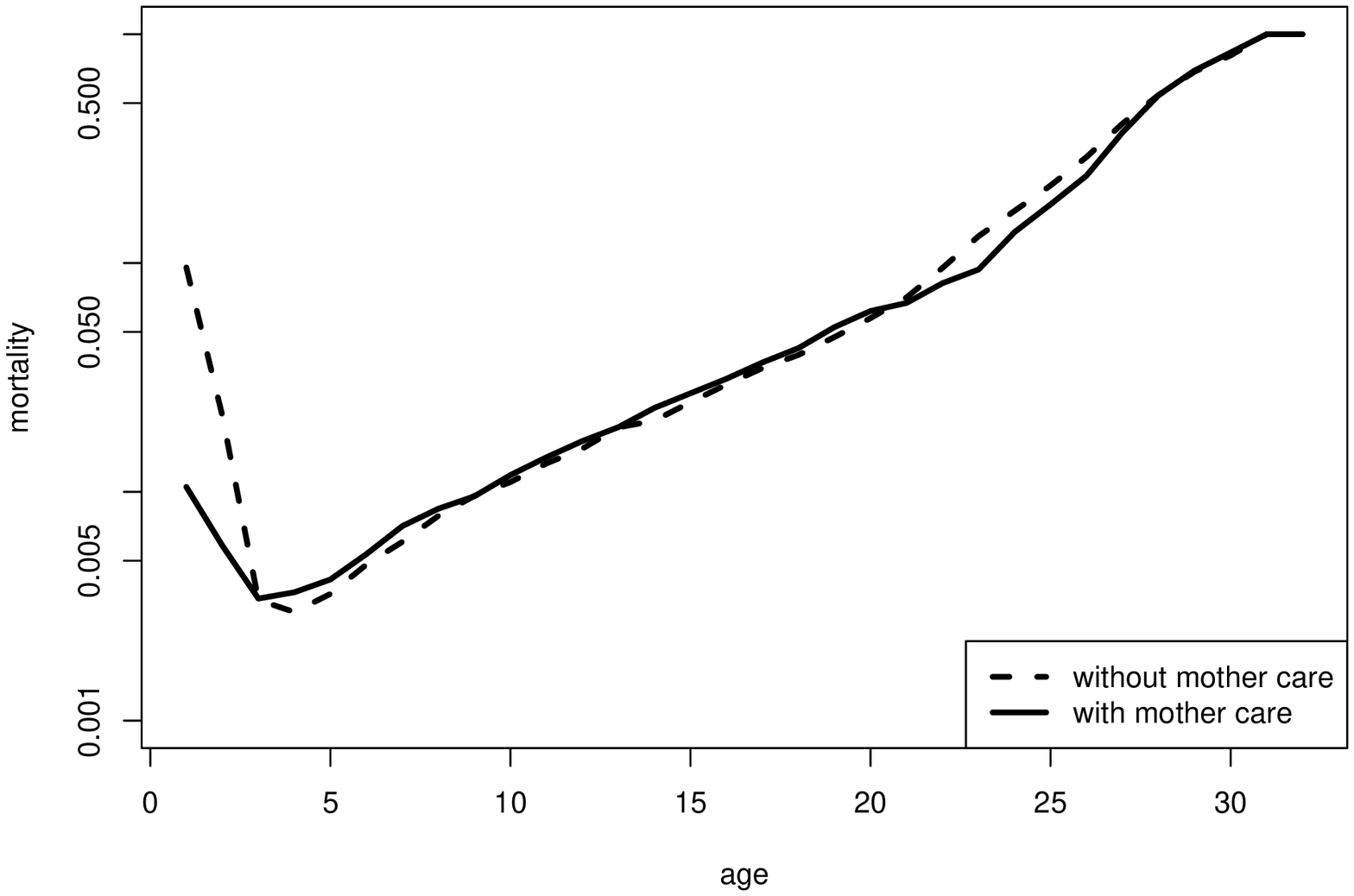}
		\caption{\label{fig13}The mortality curve for populations with and without mother care. For early periods of live the mortality of newborns is lower in case of mother care, but then it is a bit higher.}
\end{figure}

Plot shown in Fig. \ref{fig12}-\ref{fig13} indicate, that one could really expect slightly higher mortality of young individuals if they are isolated from environment influences during the very early periods of life.

\clearpage

\section{Additional risk factors}

\begin{figure}[h!]
	\centering
		\includegraphics[height=0.95\textwidth,angle=270]{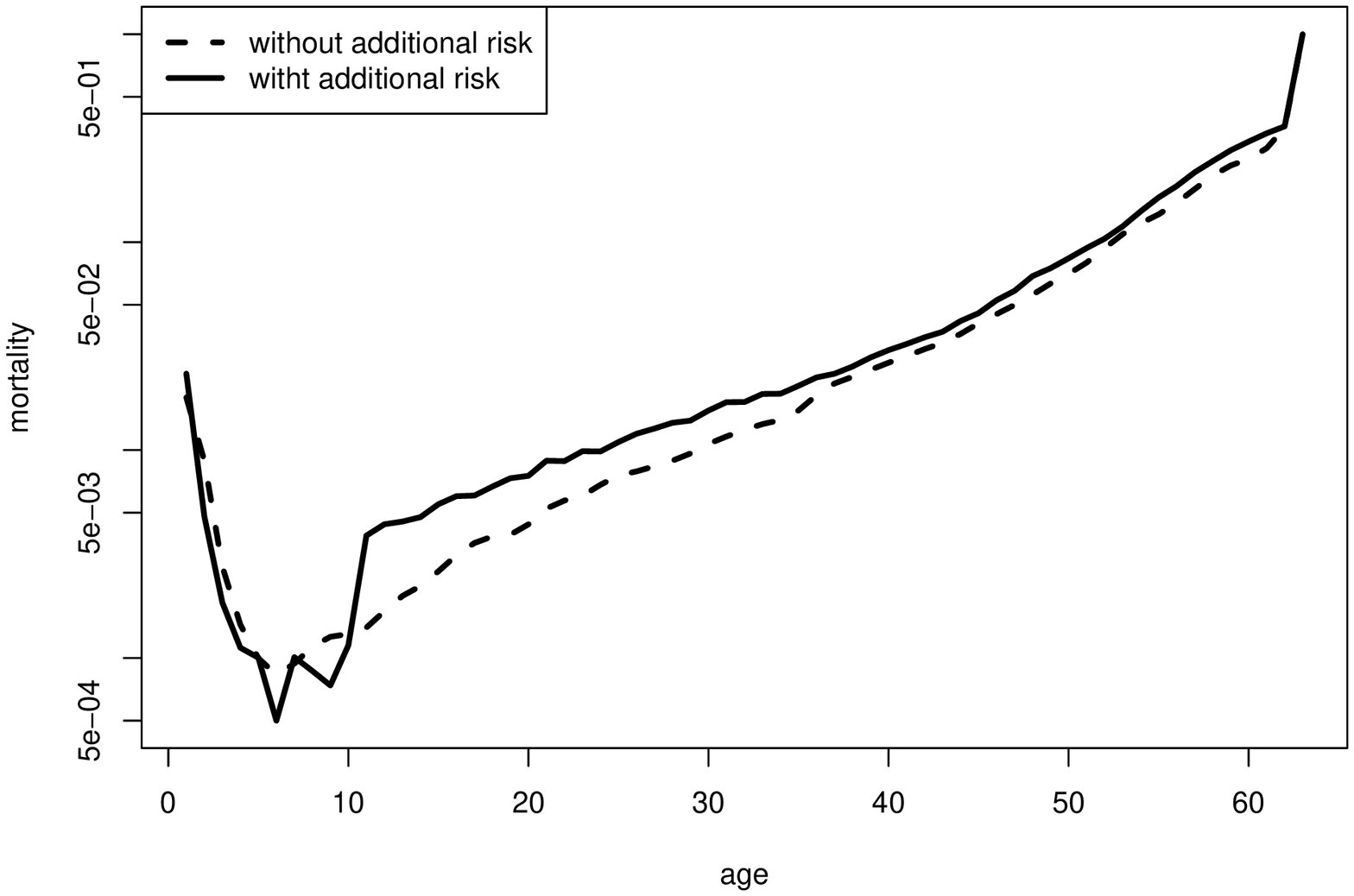}
		\caption{\label{fig14}The mortality curve for the population with additional risk of death after $H=12$ years.}
%\end{figure}
%
%\begin{figure}
%	\centering
%		\includegraphics[height=0.95\textwidth,angle=270]{figs/strukturaWiekowa.eps}
		\includegraphics[height=0.95\textwidth,angle=270]{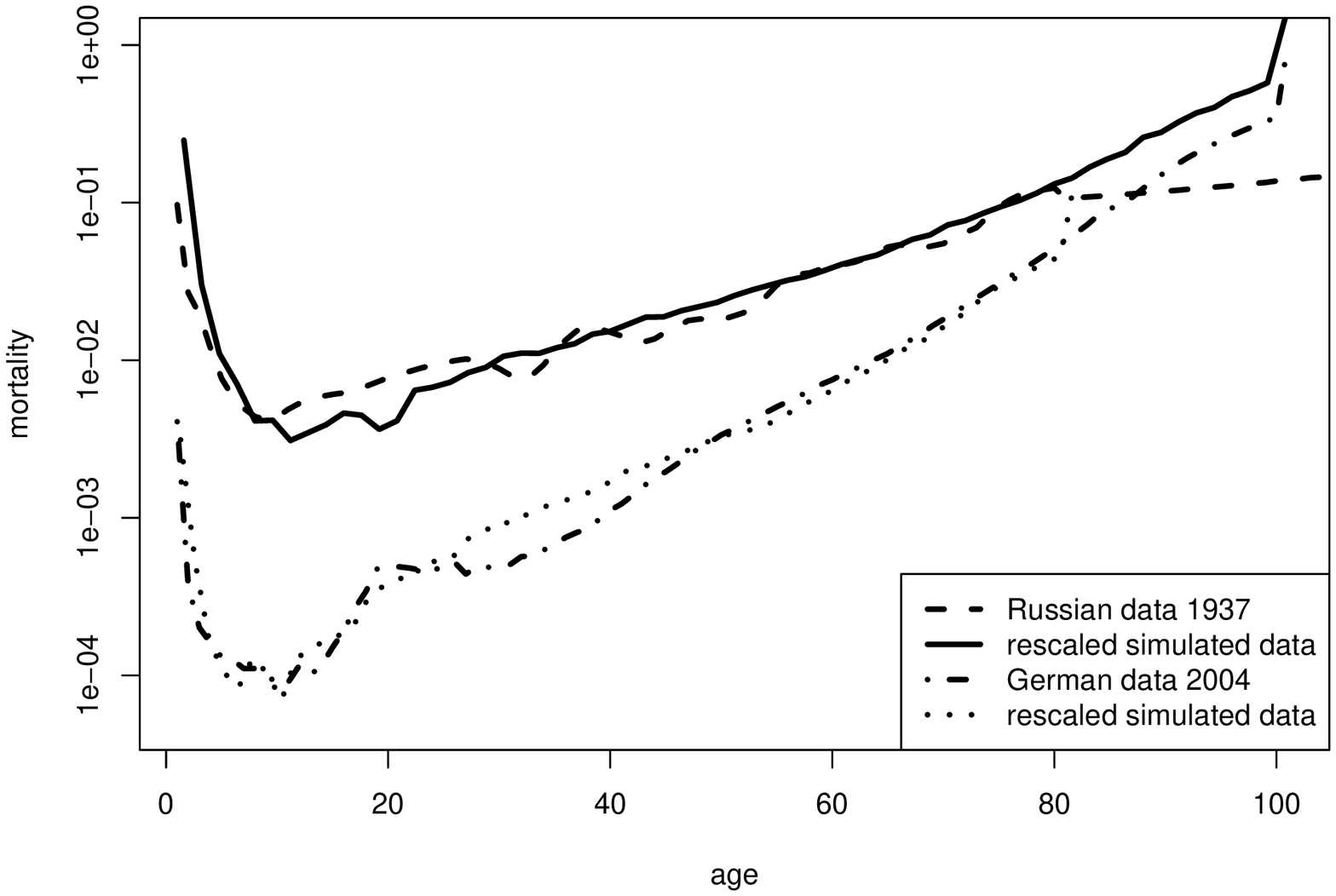}
		\caption{\label{fig15}The age structures (upper plots) and mortality curves of different human populations and populations generated in simulations with the extended noisy Penna model (lower plots).}
\end{figure}

Suppose that after age $H$ in every year an individual may die with some very small probability due to the random death (e.g. in a car accident or any other death of young people released from the parental care). In Fig. \ref{fig14} we presented results for populations, in which individuals older than $H=12$ dies in each year with probability $p=0.002$ even if their state is lover than the $F$. The mortality curve resembles the curves observed in many real human populations. This random death introduced for individuals older than 12 could be also natural increase in mortality connected with reaching the puberty age.  

In Fig. \ref{fig15} we presented the age structure of different human populations and for comparison, the mortality curves generated by simulations with properly rescaled the age axis.

In the standard Penna model, if Verhulst factor regulates the size of the population only at the level of birth, there are no random deaths later, the mortality curve is in fact of s-shape with very low death rate for the youngest (in fact no death of individuals younger than $T$) and, the mortality rate above the Gompertz law prediction for the last parts of the lifespan. In our model the shape of the mortality curves changes. Even the youngest organism may die, and the results of simulations better fit to the real mortality curve of human populations.

For the oldest part of populations  some downward deviation of mortality in comparison with the Gompertz law is observed. This deviation, called plateau, is controversial and according to Stauffer \cite{Stauffer} it is a result of imperfect demographic data available for the oldest parts of the human populations (nowadays it concerns data from the end of nineteenth century) \cite{Gavrilova}. The results of standard Penna model simulations reproduce the age distribution of the human population in its middle part after the minimum reproduction age, but not the oldest part of populations. The noisy model enables simulation and analysis of parameters which influence the mortality of the youngest individuals. Nevertheless, if the heterogeneity of populations is generated by introducing specific parameters into the Penna model, the downward deviation from the Gompertz law prediction for the oldest could be obtained in the computer simulated populations \cite{Coe,LaszkiewiczOO,biecek}. There is an open question if the heterogeneity observed in the human population is generated by noise in the demographic data or noise in the relations between humans and environment.

\clearpage
\section{Ageing and the loss of complexity}
When we consider the ageing processes, usually we are thinking about deterioration of physiological functions of organism which leads to the loss of its adaptive possibilities, improper responses to the environmental conditions and eventually to death. Those are only symptoms of ageing; the basic causes of these symptoms are somewhere deeper in the complex living system. Physiologists are aware of the complex nonlinear dynamics of organisms' functions and it is more and more accepted that diseases and ageing could be characterized by the loss of complexity \cite{Lipsitz, Goldberger}. Even in the simplest organisms, several genes and usually much more gene products interplay with each other \cite{Echols}. They are connected in the complicated metabolic networks with huge number of different feedback regulatory loops. In fact there is no single function, which is not connected with many other functions of organisms. The interplay between different functions helps the healthy organism to adapt to different environmental conditions. The loss of some connection or improper response to the demand could lead to the disease or death. Such a false response could be connected with genetic defect(s).

For the further extension of the Penna model we have used the version with noise described in the above section; 
\begin{itemize}
\item the internal ,,personal'' fluctuations can be superimposed on the fluctuations of the environment, 
\item	the sum of both fluctuations describes the health status of organism,
\item	the fluctuations higher than the assumed limit kill the individual. 
\end{itemize}
In the new version the random Gaussian fluctuations are characteristic for the normal (wild) version of a gene. Defective genes change their random fluctuations for highly correlated signal of the same energy as the previous fluctuations. The idea is biologically justified, since in many examples the highly synchronized behaviors of some activities leads to harmful effects or even deaths \cite{Goldberger, Lipsitz2} (as in case of Huntington disease).

The difference between the new proposed model and the Noisy Penna Model is in behavior of an individual state.
In this version we assume that the individual state is a composition of many small fluctuations corresponding to $L$ particular genes, thus 
\begin{equation}
P_i(t) = \sum_{j=1}^L p_{i,j}(t),
\end{equation}
where 
\begin{equation}
p_{i,j}(t) \sim \mathcal N(0,\sigma^2_A).
\end{equation}

Defects are reflected in positive correlation between the switched on defective genes. In other words, if genes $k$ and $l$ are defective and switched on then 
$$
corr(p_{i,k}, p_{i,l}) = \rho.
$$

If all $p_{i,j}(t)$ are independent, then $var(P_i(t)) = L \sigma^2_A$, but the correlations between gene functions increase the variation of personal state. The most sharp effect is if the $\rho = 1$. In this case, the variation of $L$ totally correlated signals is equal to $var(P_i(t)) = L^2 \sigma^2_A$.

\begin{figure}
	\centering
%	a) 
		\includegraphics[height=0.95\textwidth,angle=270]{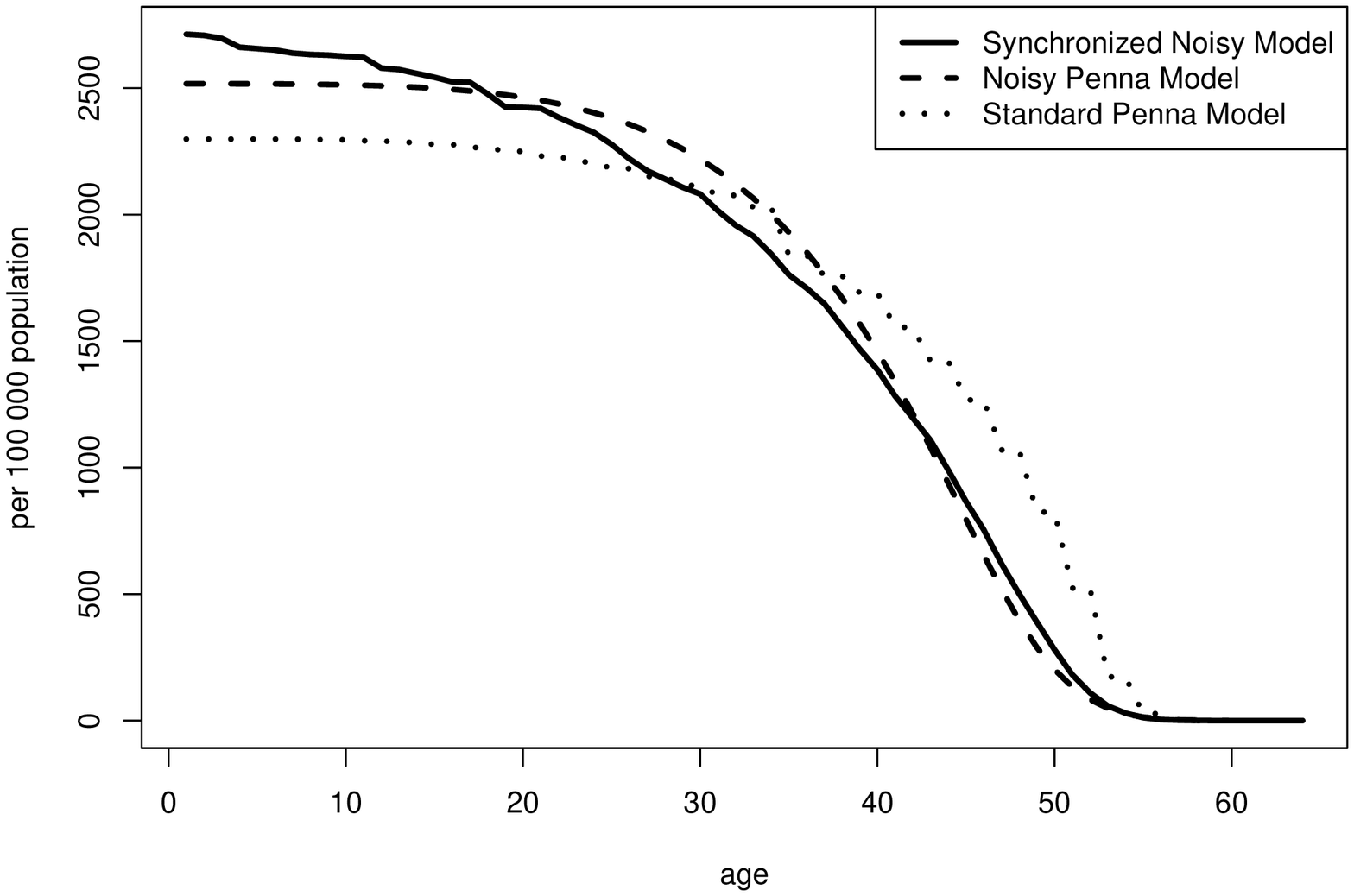}
%		\\
%	b)

		\includegraphics[height=0.95\textwidth,angle=270]{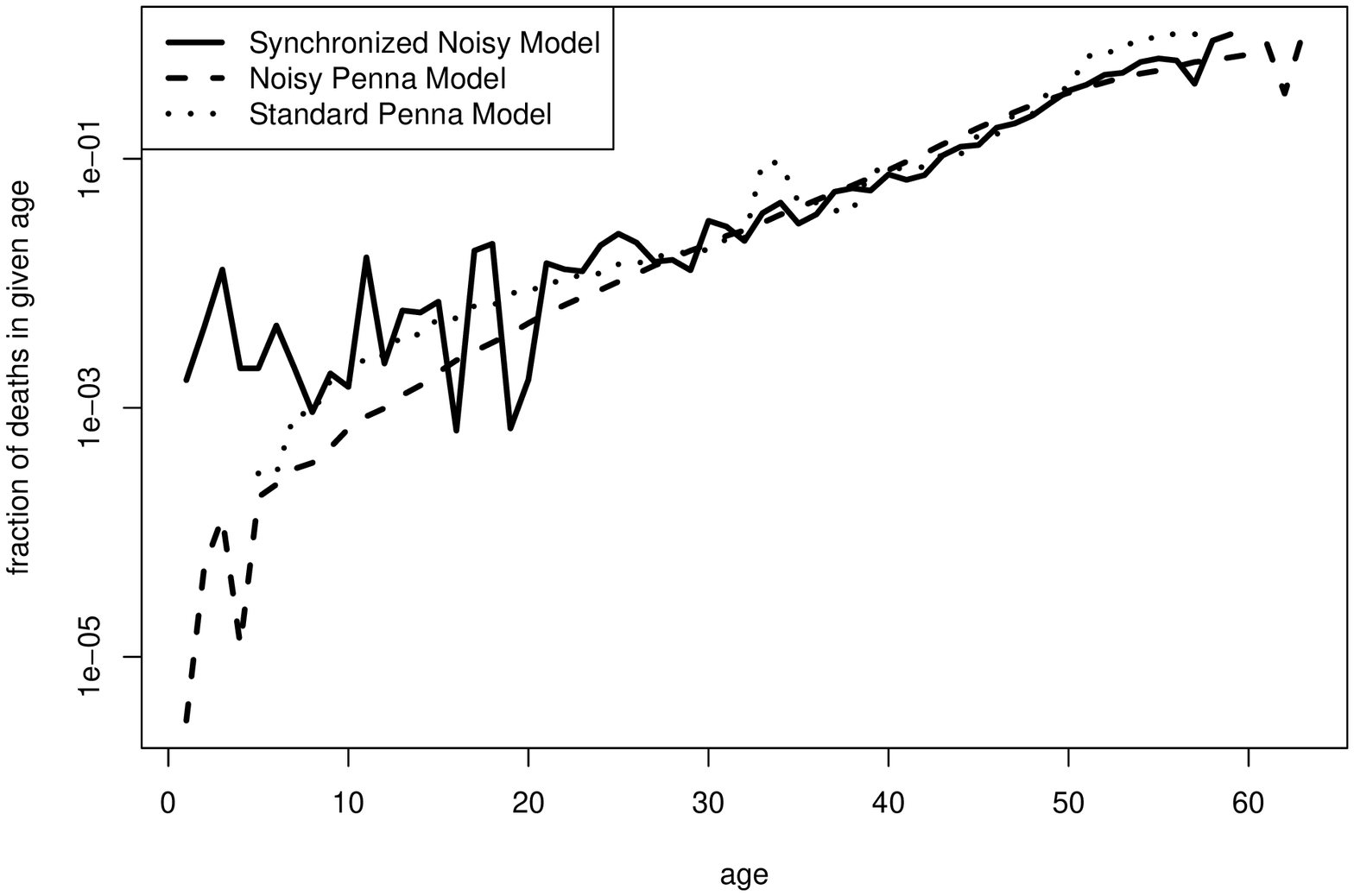}
\caption{\label{fig16}The age distribution (upper plot) and mortality curve (lower plot) for example settings of standard Penna Model (dotted line), the Noisy Penna Model (dashed line) and the proposed synchronized Penna Model (solid line). The exact values of given curves depend on the model parameters. }
\end{figure}

In Fig. \ref{fig16} we have compared the results generated by the proposed model with the results obtained with other models. The main difference is a mortality of newborns. It is much higher than in case of Noisy Penna Model and Standard Penna Model. Since the base variation is equal to composition of $L=64$ noises, it is high enough to determine the higher mortality. 

%\clearpage

\section{Why women live longer than men}
Comparing the mortality curves or age structures of men and women of many nations or ethic groups it could be clearly shown that the life expectancy of men is much lower than that of women. Especially, mortality of men at the middle age is almost 50\% higher than the mortality of women at the same age. Stauffer blamed for that many natural or social phenomena, like higher somatic mutations rate in the male bodies, stress or altruism of men - they have to fight for peace. But there could be one more reason - male are hemizygous\index{hemizygous} in respect to X chromosome - they possess only one copy of that chromosome in their genome where about 5\% of all genes (in the human genome) are localized. Women possess two copies of the X chromosome. It is very important for phenotypic expression of the defective genes located on the X chromosomes. If man has a defective gene on its copy of X chromosome, the phenotypic defective trait would be expressed. The same defective gene on one copy of X chromosome in the woman genome could be complemented by its allele located on the other copy of X chromosome. 

Assuming these differences in the genome structure of male and females Schneider et al., and Kurdziel et al., analysed the age structure of populations with such a sexual dimorphism. They found, that the differences in mortality of both sexes could be explained by the lack of the second copy of X chromosome in the male genomes. Furthermore, the model predicts higher differences in mortality in the middle ages and almost equal mortality for the oldest groups of individuals, just like in human populations. Note, that in this example authors introduced some genes expressed before the birth. The fraction of these genes was big enough to introduce so called zygotic death at the level of 60\% corresponding to estimations for humans. It is estimated that a chance for human zygote to survive until birth is of the order of 40\%. 

Thus, this simple hypothesis of the role of X chromosome has been supported by the simulations. But there is another problem for biologists. Males usually have two sex chromosomes: X and Y. There are some premises that originally these two chromosomes were homologous and in the course of evolution the Y chromosome has shrunk. Why? The Onody group used the Penna model to simulate the evolution of Y chromosome and they noticed that it accumulates mutations with higher probability than other chromosomes. They tried to explain this phenomenon by a specific genetic behaviour of the Y chromosome - it does not recombine with counterparts (in fact it has no counterparts) and it is cloned and passes only through the male genomes. Nevertheless, it was found that these properties does not explain the phenomenon of the Y chromosome shrinking. Let's start the evolution of sex chromosomes in the individuals whose genomes are composed of two pairs of chromosomes - one pair of autosomes\index{autosomes} (autosomes - all chromosomes which are not sex chromosomes) and one pair of sex chromosomes. At the beginning all chromosomes were identical and evolved exactly according to the standard diploid Penna model. The only difference was a marker glued to one copy of sex chromosomes which determined the male sex (the chromosome is called Y chromosome, from now). Thus, if a zygote was formed of two gametes - both transferring the sex chromosomes without that marker (X chromosomes) - it was a female. If a zygote got a sex chromosome with that marker - it was a male. The males could transfer the X chromosome and the Y chromosome to the gametes with an equal probability. Autosomes in all genomes and X chromosomes in the female genomes can recombine.

\begin{figure}
\centering
		\includegraphics[height=0.95\textwidth,angle=270]{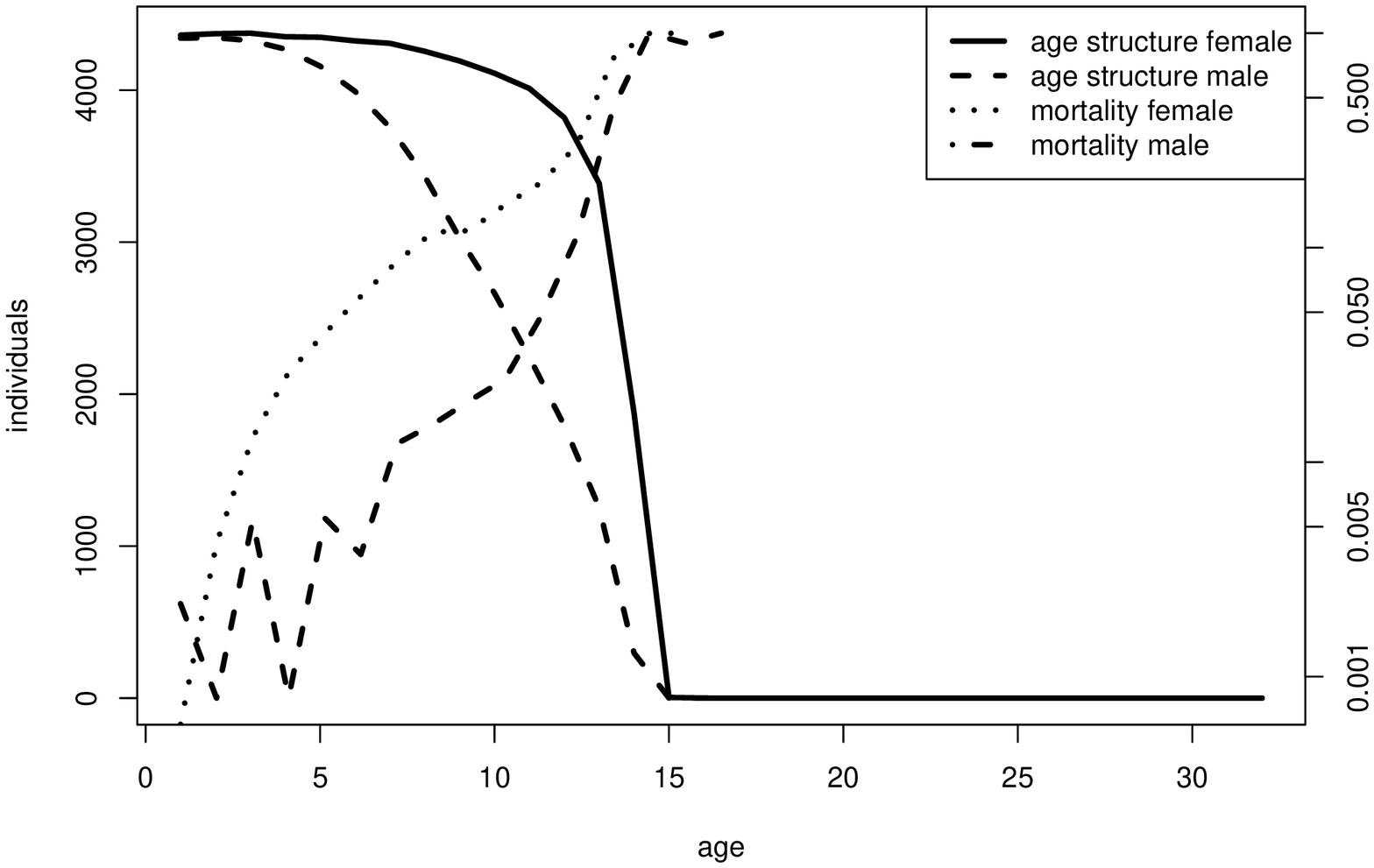}
		\includegraphics[height=0.95\textwidth,angle=270]{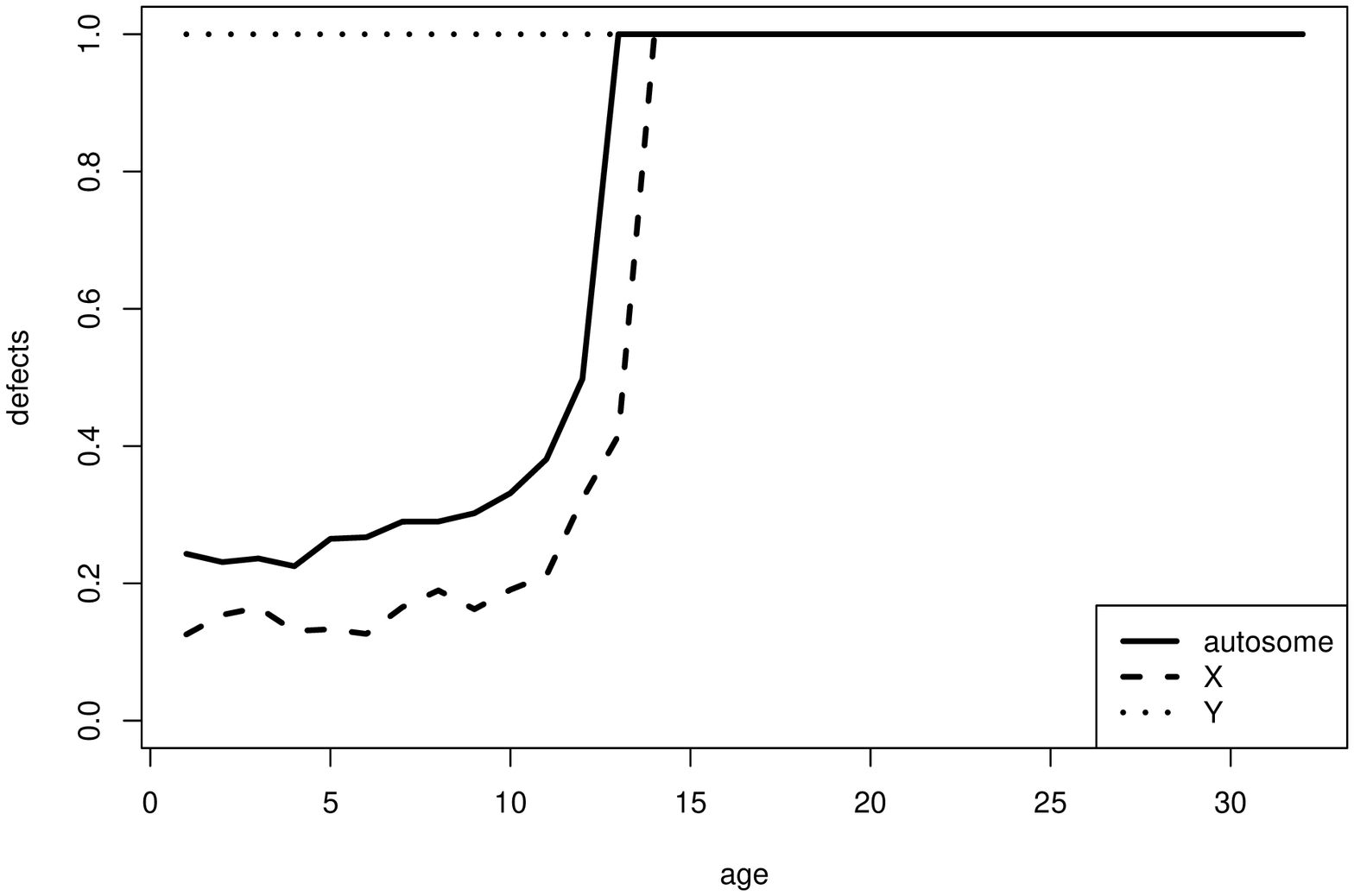}
	\caption{\label{fig17}XX/XY system - panmictic population. Age structure and mortality in populations after 20 000 MCs (the upper plots). Note the logarithmic scale (right) for mortality. Lower plots represent fraction of defective alleles in autosomes and sex chromosomes.}
\end{figure}

The result of simulations are shown in Fig. \ref{fig17}. After relatively short time of evolution, the Y chromosome accumulated defective genes and eventually the only genetic information it possessed was the determination of the male sex - the marker glued to the chromosome. The other effect of evolution was lower fraction of defective genes in the X chromosome comparing to the autosomes and the most important - the differences in the age structures and mortalities curves of males and females - females lived longer and the difference in mortality was the highest at the middle ages. The lower fraction of defective genes in X chromosomes is a result of more effective elimination of defects from these chromosomes by males where all defective genes play role of dominant defective mutations. Is the cloning the only reason for shrinking Y chromosome? In such versions of the modelling the evolution, populations are usually panmictic\index{panmictic} - females can freely choose a male partner from the whole pool of males in the reproduction age available in the population. 

Additionally, male after reproducing with one female is going back to the pool of males. Thus, one male individual can reproduce several times during the one time unit while each female can reproduce only once in the time step. This property of panmictic population has been changed - male individual after reproduction could not go back to the pool of males in the population, it has to be faithful to its female for whole life. This simple change dramatically changed the fate of Y chromosome. The Y chromosome does not shrink any more. 
One can conclude that if women are not promiscuous and do not seduce the married men than Y chromosome could preserve its all genetic information and men could live as long as women.

%\clearpage

It is still not the whole story of the sex chromosome evolution. In the above versions of the model it has been assumed that the recombination between X and Y chromosomes was switched off from the beginning of the evolution. Biologists are not sure how and when the recombination between these chromosomes was switched off. But if recombination between X and Y chromosome is advantageous then any modification of this process decreasing the crossover rate should be eliminated by selection. To keep the lower recombination rate as an adaptation, one has to assume that it is profitable. Thus, it should be possible to observe self organization of recombination rate during the simulation of the population evolution. At the beginning of simulations, one pair of autosomes, X - X pair and X - Y pair of sex chromosomes recombined with an equal frequency 0.5. The newborns inherited the recombination rate as an average of both parents but this inherited value of recombination rate could be changed in their genomes by +/- 0.01 independently for each pair of chromosomes. During the evolution, the recombination rate between autosomes and between X chromosomes increased and fluctuated around one crossover event per gamete production while the recombination between X and Y chromosome was switched off very fast (it stayed at the low level corresponding to the fluctuations caused by the increment of its changes).

\textbf{The role of the crossover rate for the strategies of evolution}
\index{crossover}
In the standard diploid Penna model recombination\index{recombination} between the bitstrings mimics the crossover and usually it is assumed that the frequency of crossover is 1 per bitstrings' pair during the gamete production. There are some other possibilities of the implementation of crossover, very often used in other models. One of them assumes that during the gamete production one allele is randomly drawn from each pair of loci from the bitstrings. This method is not biologically justified because it assumes that all genes are inherited independently - that there are no genetic linkages between genes, which is obviously not true. Here we would like to show how unexpected results can be obtained during modelling the population evolution and how complicated and difficult are biological interpretations of such results. 

\begin{figure}[h!]
		\includegraphics[height=0.95\textwidth,angle=270]{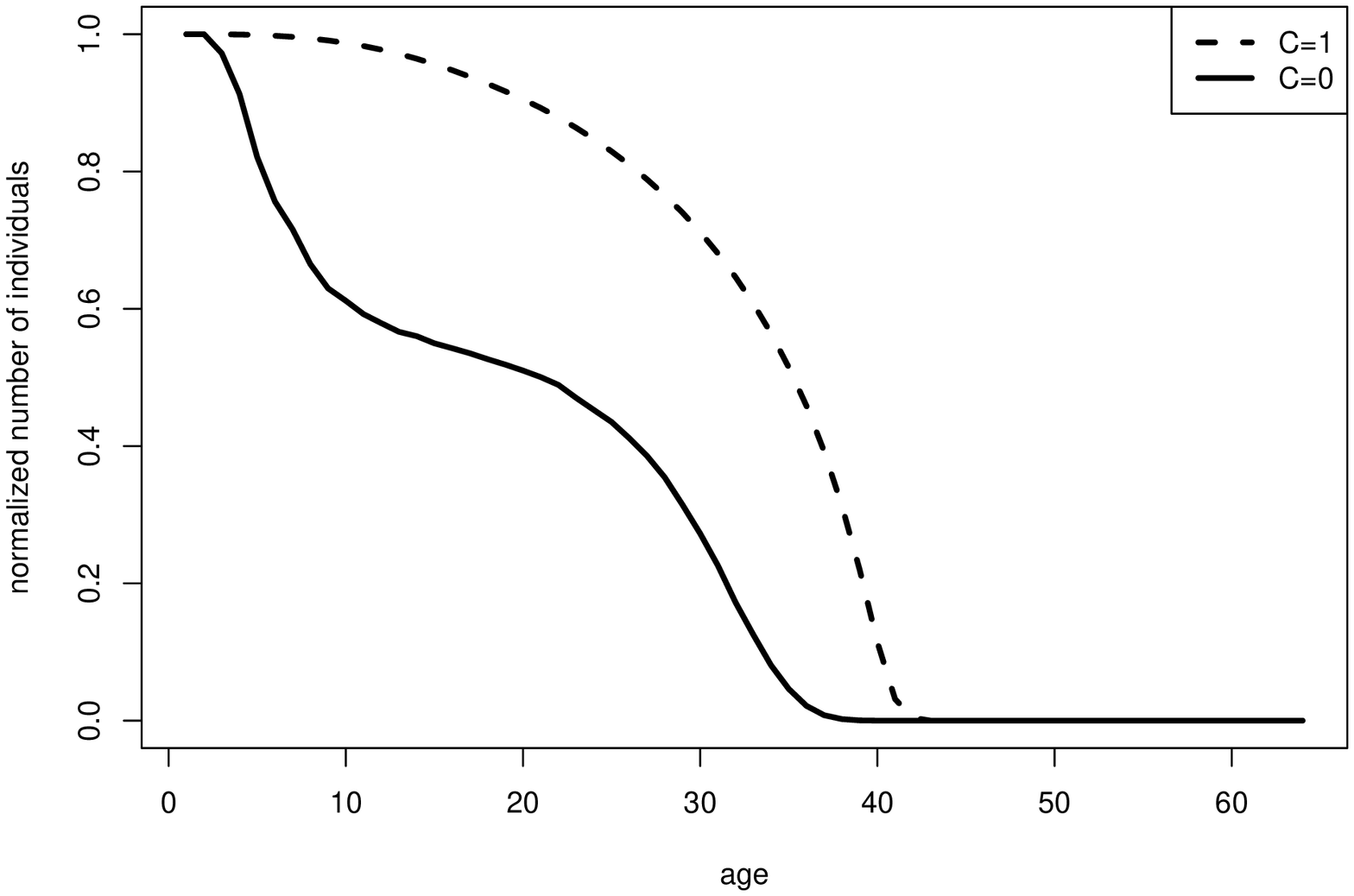}
		\includegraphics[height=0.95\textwidth,angle=270]{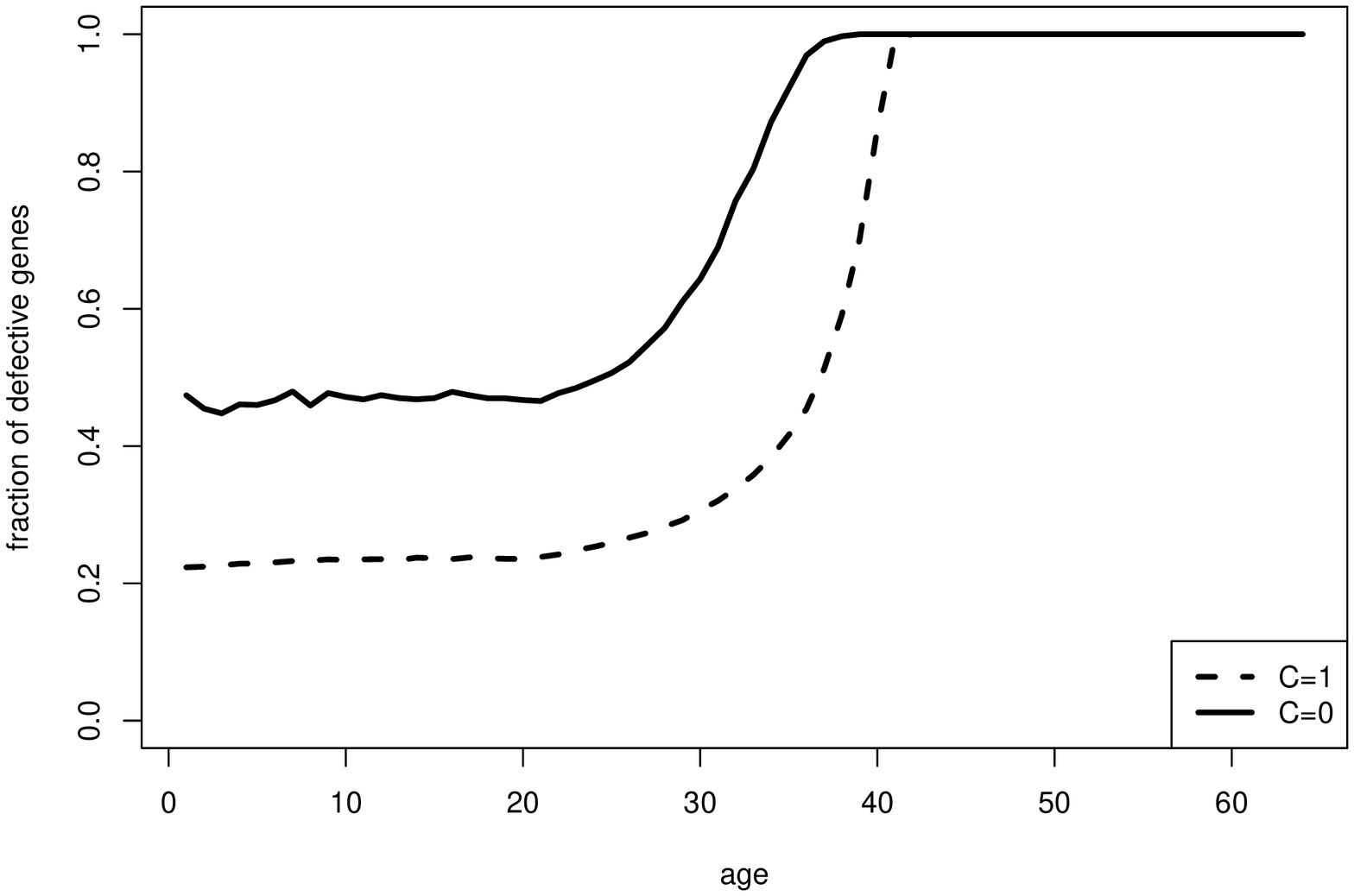}
	\caption{\label{fig18}Results for different rates of crossover.	 Higher crossover rate results in lower fraction of defective genes and higher fraction of individuals in reproduction age.}
\end{figure}

We have compared the results of two different simulations: with one recombination event between the bitstrings (haplotypes) and without recombination. In the second version, one randomly drawn haplotype after mutation was considered as gamete. The results are shown in Fig. \ref{fig18}. 

%\clearpage

The populations evolving without recombination are smaller and they have very high level of defective genes even in the part of the genomes expressed before the reproduction age. The frequency of defects reaches 0.5. The minimum reproduction age was set for 20, thus, assuming the random distribution of defects at this part of genome and that all mutations are recessive it is easy to calculate the probability of expression the phenotypic trait of a single locus. It equals 0.25. Thus, on average, five defects should be expressed before the reproduction age, substantially decreasing the reproduction potential of population, in fact population should die. Nevertheless, populations survive for very long time and no signs of extinction had been observed. This puzzle will be a subject of the next section which describes the two strategies of genomes' evolution. 

\clearpage
 
\bibliographystyle{ws-rv-van}

\section*{Acknowledgements}
Calculations have been carried out in Wroc\l{}aw Centre for Networking and Supercomputing (http://www.wcss.wroc.pl), grant \#102.

%\printindex[aindx]                 % to print author index
\printindex                         % to print subject index
\end{document}